%% file: main.tex
\input{meta/header}
\input{meta/comments}

\input{meta/acronym}
\input{meta/templates}
\input{meta/mathutils}

\AtBeginDocument{%
  }

\setcopyright{acmlicensed}
\copyrightyear{2018}
\acmYear{2018}
\acmDOI{XXXXXXX.XXXXXXX}
\acmConference[Preprint]{XXX}{December 14--18,
  2026}{Tarragona, Spain}
\acmISBN{978-1-4503-XXXX-X/2018/06}

\setcopyright{none}

\begin{document}


\title[Efficient and Robust Online Learning to Rank in Decentralized Systems]{Efficient and Robust Online Learning to Rank\\in Decentralized Systems}

\subtitle{Research}  


\author{Marcel Gregoriadis}
\orcid{0000-0001-5094-0111}
\affiliation{%
  \institution{TU Delft}
  \city{Delft}
  \country{Netherlands}
}
\email{m.gregoriadis@tudelft.nl}

\author{Martijn de Vos}
\orcid{0000-0003-4157-4847}
\affiliation{%
  \institution{EPFL}
  \city{Lausanne}
  \country{Switzerland}
}
\email{martijn.devos@epfl.ch}

\author{Sayan Biswas}
\orcid{0000-0002-2115-1495}
\affiliation{%
  \institution{EPFL}
  \city{Lausanne}
  \country{Switzerland}
}
\email{sayan.biswas@epfl.ch}

\author{Anne-Marie Kermarrec}
\orcid{0000-0001-8187-724X}
\affiliation{%
  \institution{EPFL}
  \city{Lausanne}
  \country{Switzerland}
}
\email{anne-marie.kermarrec@epfl.ch}

\author{Johan Pouwelse}
\orcid{0000-0002-9882-1506}
\affiliation{%
  \institution{TU Delft}
  \city{Delft}
  \country{Netherlands}
}
\email{j.a.pouwelse@tudelft.nl}

\renewcommand{\shortauthors}{Gregoriadis et al.}

\input{sections/abstract}




\maketitle
\pagestyle{preprint}



\input{sections/introduction}
\input{sections/preliminaries}
\input{sections/doltr}
\input{sections/theory}
\input{sections/experimental_setup}
\input{sections/experiments}

\input{sections/related_work}
\input{sections/conclusion}
\bibliographystyle{ACM-Reference-Format}
\bibliography{ref}

\appendix

\input{appendix/notations}
\input{appendix/convergence_proof}
\input{appendix/robustness_plots}

\end{document}

%% file: meta/header.tex
\PassOptionsToPackage{table}{xcolor}
\documentclass[sigconf,9pt]{acmart}

\renewcommand\footnotetextcopyrightpermission[1]{}  
\usepackage{fancyhdr}
\fancypagestyle{preprint}{%
  \fancyhf{}%
  \fancyhead[L]{Preprint}%
  \fancyhead[R]{Gregoriadis et al.}%
}

\usepackage{multirow}
\usepackage{tikz}
\usepackage{amsmath}
\usepackage{siunitx}
\usepackage[abbreviations]{foreign}
\usepackage{xspace}
\usepackage[inline]{enumitem}
\usepackage[ruled,vlined,linesnumbered]{algorithm2e}
\SetKwComment{Cmt}{\color{gray}\sffamily\small // }{}
\definecolor{orange}{rgb}{1,0.5,0}
\definecolor{Green}{rgb}{0,0.5,0}
\definecolor{Blue}{rgb}{0,0,1}
\usepackage{svg}

\usepackage{booktabs, makecell, pifont} 
\usepackage{cleveref}
\usepackage{bm}
\usepackage{url}
\usepackage{subcaption}
\usepackage{tabularx}

\usepackage{cuted}
\usepackage{caption}

\definecolor{pink}{RGB}{173, 38, 30}
\definecolor{green}{RGB}{81, 166, 56}
\definecolor{stepbg}{RGB}{0, 0, 0}


%% file: meta/comments.tex
\usepackage{ifthen}
\newboolean{showcomments}
\setboolean{showcomments}{false}
\ifthenelse{\boolean{showcomments}}
{ \newcommand{\mynote}[3]{
		\fbox{\bfseries\sffamily\scriptsize#1}
		{\small$\blacktriangleright$\textsf{\emph{\color{#3}{#2}}}$\blacktriangleleft$}}
	\newcommand{\zzz}[1]{{\setlength{\fboxsep}{2pt}\fcolorbox{black}{yellow}{\textsf{\emph{#1}}}}\xspace}}
{ \newcommand{\mynote}[3]{}
	\newcommand{\zzz}[1]{}}

\definecolor{maroon}{rgb}{0.5, 0.0, 0.0}

\newcommand{\amk}[1]{\mynote{AMK}{#1}{red}}
\newcommand{\mg}[1]{\mynote{Marcel}{#1}{orange}}

\newcommand{\mv}[1]{\mynote{Martijn}{#1}{brown}}

\newcommand{\TODO}[1]{\zzz{TODO: #1}}

%% file: meta/acronym.tex
\usepackage{acronym}
\acrodef{ML}{machine learning}
\acrodef{RCA}{root cause analysis}
\acrodef{IaC}{infrastructure as code}
\acrodef{LLM}{large language model}
\acrodef{SERP}{search engine result page}
\acrodef{LTR}{Learning to Rank}
\acrodef{OLTR}{Online Learning to Rank}
\acrodef{FOLTR}{Federated Online Learning to Rank}
\acrodef{DL}{decentralized learning}
\acrodef{FL}{federated learning}
\acrodef{GL}{gossip learning}
\acrodef{PDGD}{Pairwise Differentiable Gradient Descent}
\acrodef{D-PSGD}{Decentralized Parallel Stochastic Gradient Descent}
\acrodef{RPS}{Random Peer Sampler}
\acrodef{SDBN}{Simplified Dynamic Bayesian Network}

%% file: meta/templates.tex
\newcommand{\sys}{\textsc{RankGuard}\xspace}

\newcommand{\flip}{\textsc{Flip}\xspace}
\newcommand{\lie}{\textsc{LIE}\xspace}
\newcommand{\ipm}{\textsc{IPM}\xspace}
\newcommand{\adapt}{\textsc{Adapt}\xspace}
\newcommand{\fltrust}{\textsc{FLTrust}\xspace}
\newcommand{\zenops}{\textsc{ZenoPS}\xspace}
\newcommand{\elloc}{\textsc{EL-Local}\xspace}

%% file: meta/mathutils.tex
\usepackage{physics}
\usepackage{thm-restate}
\usepackage{bbm}
\newtheorem{theorem}{Theorem}

\newtheorem{remark}[theorem]{Remark}
\newtheorem{assumption}[theorem]{Assumption}

\allowdisplaybreaks

%% file: sections/abstract.tex
\begin{abstract}
In Online Learning to Rank (OLTR), ranking models are trained directly from live user interactions, but existing systems rely on a trusted central server to collect and process these interactions.
This leaves operators free to introduce biases that conflict with user interests.
Decentralized learning offers an attractive alternative, allowing users to collaboratively train a shared ranking model by exchanging model updates directly with one another, without any central authority.
In such settings, however, malicious nodes can send poisoned model updates that degrade the ranking quality of honest nodes.
We introduce \sys, a decentralized OLTR framework in which users collaboratively train ranking models and exchange model updates directly with other nodes.
\sys defends against poisoning attacks by carefully evaluating incoming models against the user's own private click history, corrected for position bias.
An incoming model is only aggregated if it better explains the user's past interactions than the current local model, making it fundamentally hard for malicious nodes to craft updates that pass this test without also genuinely helping the user. 
We derive a theoretical convergence guarantee of \sys. To the best of our knowledge, this is the first formal convergence analysis of a decentralized OLTR algorithm.
We evaluate \sys against four poisoning attacks, including a powerful adaptive attack, using four standard benchmarks and three click models.
\sys outperforms all baselines in most settings while being up to $62\times$ more efficient than its closest competitors.
\end{abstract}

%% file: sections/introduction.tex
\section{Introduction}

Search rankings shape what people see and believe~\cite{epstein2015search}.
Most users never scroll past the first few search results~\cite{pan2007google}.
Therefore, a nudge to these results has the power to steer consumer purchases~\cite{ghose2014examining,ursu2018power}, influence health decisions~\cite{pogacar2017positive}, shape beliefs about contested topics~\cite{draws2021not}, and even sway election outcomes~\cite{epstein2015search,epstein2017suppressing,eu2024tiktok}.
Today, this power is concentrated in the hands of a few search engine operators, such as Google and Baidu.
This allows them to introduce biases that conflict with user objectives, \eg, for commercial gain~\cite{farronato2023self}. 
The factors that drive these ranking decisions remain opaque to the user.
This motivates an alternative system that does not rely on a centralized authority for ranking search results.

Modern ranking systems are trained from user interactions with \acp{SERP}.
A \ac{SERP} presents the user with an ordered list of documents matched to their query and ranked by predicted relevance.
Within this domain, \Ac{OLTR} is a prominent paradigm for training ranking models directly from live user interactions~\cite{hofmann2013balancing,oosterhuis2018differentiable}.
\Ac{OLTR} treats user interactions (typically, \emph{clicks}) on these documents as implicit relevance signals to iteratively update the ranker, typically via \ac{ML}.
Most \ac{OLTR} systems rely on a central server to aggregate click feedback and serve rankings.
\Ac{FOLTR}~\cite{kharitonov2019federated,wang2021effective} keeps click data on the user's device and shares only model updates, but still relies on a trusted central server to aggregate those updates and distribute the resulting ranker.

\Ac{DL} sidesteps the need for a central server entirely~\cite{ormandi2013gossip,de2023epidemic}. 
With \ac{DL}, nodes train locally and exchange model updates directly with one another via peer-to-peer protocols, aggregating received updates into their own models. 
While this removes dependence on a central server, it exposes nodes to attacks from any other node in the network~\cite{el2020genuinely}.
In particular, adversarial nodes can send maliciously crafted model updates that, once aggregated, degrade other users' rankers.
This attack is known as an \emph{untargeted poisoning attack}, and has primarily been studied in the context of \ac{FL}~\cite{fang2020local}.
Yet, existing defenses were designed for federated settings with a central server and do not transfer cleanly to decentralized \ac{OLTR}, where no such server exists and nodes must independently assess the trustworthiness of received model updates.
Therefore, the problem of robust \ac{OLTR} in decentralized settings remains unexplored. 
To the best of our knowledge, this work is the first to address this issue.

\begin{figure}[t]
    \centering
    \includegraphics[width=\linewidth]{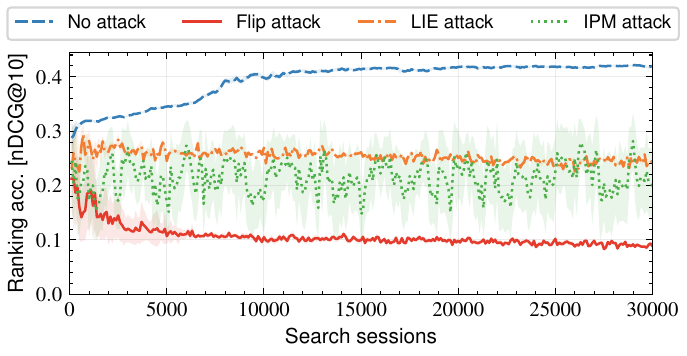}
    \caption{Ranking accuracy of honest nodes as users interact with the system 
    with 100 nodes of which 20 send malicious model updates. 
    }
    \label{fig:motivation}
\end{figure}

To demonstrate the impact of malicious nodes on the ranking accuracy of honest nodes, we conduct an experiment.
We set up a 100-node network with 20\,\% of these nodes performing three standard poisoning attacks (\flip, \lie and \ipm).
After each session, a node updates its model and shares it with 7 random nodes, who aggregate it locally.
The full experimental setup is provided in \Cref{sec:exp_setup}.
\Cref{fig:motivation} shows the average ranking accuracy of honest nodes over time in terms of nDCG@10, a standard accuracy metric in \ac{OLTR} that scores how well the top-10 results match ideal relevance ordering (1 = perfect, 0 = worst).
Under all three attacks, malicious nodes significantly degrade ranking accuracy for honest nodes.
Notably, after \num{30000} search sessions, average accuracy drops from 0.42 to 0.09 under the \flip attack, rendering the ranking system unusable.
This experiment demonstrates that ranking manipulation is a serious concern in decentralized \ac{OLTR}.

Decentralized platforms that would naturally benefit from robust decentralized \ac{OLTR} already exist.
For example, \textsc{Filecoin} must rank storage providers across a distributed network by price, reliability, and latency~\cite{guidi2022evaluating}, and \textsc{DTube} must surface relevant videos without a central server~\cite{doan2020measuring}.
For such systems, a centralized ranking component would reintroduce the very bottleneck and trust assumptions that decentralization is designed to eliminate, making a robust decentralized \ac{OLTR} mechanism essential.

%
%

We address this gap and introduce \sys, a robust aggregation mechanism designed specifically for \ac{OLTR} in decentralized settings.
The key idea is to turn each user's own interaction history into a reference: when a model update is received, we evaluate how well it explains the user's past interactions, corrected for position bias, and compare this against the user's current local model.
This comparison is formalized as a statistical test whose outcome determines how much of the received model is integrated in the current local model.
A model that genuinely improves ranking quality will better explain the user's past clicks and be up-weighted, whereas malicious model updates, trained against the user's interests, will score poorly and be discarded.
This makes it fundamentally challenging for malicious nodes to craft model updates that pass this test without also improving the ranking models of honest users.


We implement \sys and conduct an extensive experimental evaluation across four standard learning-to-rank benchmarks (WEB30K, MQ2007, Yahoo, and Istella), three user behavior models, and with four poisoning attacks of increasing sophistication, including a challenging adaptive attack that assumes oracle knowledge, tailored specifically to \sys.
Our results show that \sys outperforms all baselines in the majority of settings.
We demonstrate that \sys is 62$\times$ more computationally efficient than its closest competitors, making it practical for deployment in large-scale decentralized systems.
\amk{why? would be good to give the intuition} \amk{also which is the closest competitor?}
Thus, \sys lays the groundwork for fully decentralized search mechanisms that are both robust to manipulation and free from centralized control.

\textbf{Contributions.} This work makes the following contributions:

\begin{enumerate}
    \item We introduce \sys, an efficient mechanism for robust decentralized \ac{OLTR}. \sys thwarts poisoning attacks by having nodes carefully judge incoming model updates against their own session history (\Cref{sec:rankguard}).
    \item We provide the first formal convergence analysis of a decentralized \ac{OLTR}. Specifically, we introduce a novel theoretical approach to account for the effect of the dynamic aggregation weights determined by each node to quality-gate the models they receive to derive the theoretical convergence bounds for \sys (\Cref{sec:theory}).
    \item We extensively evaluate \sys across four learning-to-rank benchmarks, three click models, and four poisoning attacks, demonstrating that it outperforms all baselines in the majority of settings while being up to 62$\times$ more efficient than its closest competitors (\Cref{sec:exp_setup} and \Cref{sec:experiments}).
\end{enumerate}






%% file: sections/preliminaries.tex
\section{Background and Problem Description}

This section introduces the necessary background for \sys. 
We begin with an overview of \ac{OLTR} and then describe \ac{PDGD}, a state-of-the-art \ac{OLTR} algorithm underlying \sys.
An overview of the notation used in this paper is provided in Appendix~\ref{app:notation}.

\subsection{(Online) Learning to Rank}\label{sec:oltr}

\textbf{\Acf{LTR}} is the task of training an \ac{ML} model to order a set of documents by relevance to a query~\cite{liu2009learning}. 
Formally, given a query $q$ and a candidate document set $D$, a ranking model $f_\theta$ assigns a score $f_\theta(\bm{d})$ to each document $\bm{d} \in D$.
Such a ranking model typically operates on a feature vector representing the query-document pair, capturing relevance signals such as BM25 score or TF-IDF.
Documents are then presented to the user in decreasing score order.
Training such a model traditionally requires editorial relevance judgements, in which human assessors assign relevance labels to query-document pairs.
These labels serve as supervision for pointwise, pairwise, or listwise loss functions~\cite{burges2005learning, cao2007learning}. 
Even though manual editorial annotation is effective, it is very labor-intensive.
Moreover, it does not account for personal preferences, as well as concept drift.

\textbf{\Acf{OLTR}} addresses these limitations by learning relevance rankings directly from implicit user feedback (most commonly clicks) collected during live interaction with the system~\cite{yue2009interactively, oosterhuis2018differentiable}. 
An \ac{OLTR} system presents ranked lists to users from a provided query, observes which documents are clicked, and updates the model accordingly.
This makes \ac{OLTR} particularly attractive in settings where annotation is unavailable or where relevance preferences are personal and evolve over time.
A central challenge in \ac{OLTR} is \emph{position bias}: users are more likely to examine and click documents ranked near the top, regardless of their true relevance.
Therefore, naive treatment of clicks as relevance labels introduces systematic bias into the model.
Counterfactual approaches~\cite{joachims2017unbiased} and pairwise methods such as Dueling Bandit Gradient Descent~\cite{yue2009interactively} and \ac{PDGD}~\cite{oosterhuis2018differentiable} address this by reweighting clicks or sidestepping the bias through interleaving.

\subsection{Pairwise Differentiable Gradient Descent}\label{sec:pdgd}
Our work builds on \ac{PDGD}~\cite{oosterhuis2018differentiable}, which provides a principled approach to debiasing user clicks and represents the state of the art in \ac{OLTR}.
It consists of two key ideas: probabilistic ranking from a sampling distribution, and a pairwise optimization strategy using a position-bias reweighting function applied to pairwise click preferences.

\ac{PDGD} considers a pointwise scoring function $f_\theta(\bm{d}): \mathbb{R}^d\mapsto [0,1]$, where $\bm{d}\in \mathbb{R}^d$ represents the features of a query-document pair in a $d$-dimensional space.
The learning goal is to find the parameters $\theta$ for which the ranker displays an optimal ranking of documents in $D$.
A higher score $f_\theta(\bm{d})$ indicates higher predicted relevance of $\bm{d}\in D$.
Rather than sorting documents directly by score, \ac{PDGD} uses the scores to define a probability distribution over documents using the Plackett-Luce model
$P(\bm{d} \mid D)
= \frac{\exp\!\left(f_\theta(\bm{d})\right)}
{\sum_{\bm{d'} \in D} \exp\!\left(f_\theta(\bm{d'})\right)}$.
A ranking $R$ of length $k$ is then created by sampling from the distribution $k$ times. 
This allows us to model the probability of a ranking:
\begin{equation}\label{eq:pdgdSampling}
    P(R \mid D) = \prod_{i=1}^{k} P\!\left(\bm{d_i} \mid D \setminus \{\bm{d_1}, \ldots, \bm{d_{i-1}}\}\right)
\end{equation}
The probabilistic sampling of a ranking serves two purposes:
\begin{enumerate*} 
\item it introduces controlled exploration, giving lower-scored documents a chance to be observed, \item and it enables counterfactual reasoning about alternative orderings, which underpins the position-bias correction described next.
\end{enumerate*}

The user may click one or multiple documents in the displayed ranking.
\ac{PDGD} assumes that clicks indicate a preference for the respective documents over the non-clicked documents in the list, albeit influenced by position bias.
As a caveat, non-clicked documents are only considered if they were also examined.
A simple but common strategy is to assume all documents up to the clicked document plus the subsequent document (if available) were examined.
This strategy is based on eye-tracking studies~\cite{joachims2005accurately}.
Specifically, \ac{PDGD} derives a set of pairwise preferences $\bm{d_i} >_c \bm{d_j} \in \mathcal{P}$ between each clicked document $\bm{d_i}$ and non-clicked but examined document $\bm{d_j}$ in the result list.
Each pairwise preference provides a biased signal.
To quantify this bias, \ac{PDGD} proposes to contemplate the probability of the swapped ranking $R^*$.
The swapped ranking is the counterfactual: the same ranking with $\bm{d_i}$ and $\bm{d_j}$ in opposite positions. Comparing $P(R \mid D)$ to $P(R^* \mid D)$ reveals how much the ranker, rather than chance, drove $d_i$ above $d_j$. When $P(R \mid D) \gg P(R^* \mid D)$, the ranker dictated the order and a click on $d_i$ largely reflects position bias; when the two are comparable, the user clicked $d_i$ against the ranker's pull, \ie, the user's click carries genuine preference signal.
This allows us to quantify the position bias in each preference pair $\bm{d_i} >_c \bm{d_j}$ as:
\begin{equation}\label{eq:rho}
    \rho(\bm{d_i}, \bm{d_j}, R, D) = \frac{P\!\left(R^*(\bm{d_i}, \bm{d_j}, R) \mid D\right)}{P(R \mid D) + P\!\left(R^*(\bm{d_i}, \bm{d_j}, R) \mid D\right)}.
\end{equation}
\ac{PDGD} treats the $\rho$-weighted sum of pair-preference probabilities as a utility to maximize, and ascends its gradient:
\begin{equation}
\nabla f_\theta(\cdot) \approx \sum_{\bm{d_i} >_c \bm{d_j} \in \mathcal{P}} \rho(\bm{d_i}, \bm{d_j}, R, D)\cdot \nabla P(\bm{d_i} \succ \bm{d_j}).
\end{equation}
\citet{oosterhuis2018differentiable} proved that the resulting gradient is unbiased with respect to the user's true pairwise preferences.
\ac{PDGD} therefore yields a principled \ac{OLTR} update from each search session, and forms the local training step in decentralized \ac{OLTR}.

\subsection{Decentralized Learning}

\Acf{DL} is a collaborative \ac{ML} paradigm in which a set of nodes jointly train a shared \ac{ML} model without coordination by a central server~\cite{beltran2023decentralized}.
In \ac{DL}, training data never leaves the device; only model updates are exchanged, offering privacy benefits over centralized approaches.
This property is particularly valuable in the \ac{OLTR} setting, where user queries and click behavior can reveal sensitive information such as health concerns, financial interests, or political beliefs.
In most \ac{DL} algorithms, nodes maintain a local model, train it on their own data held locally, and periodically exchange model updates with a subset of other nodes in a communication graph.
Upon receiving model updates, nodes aggregate them into their local model, typically via a weighted average, and this process repeats until convergence.
A standard protocol in this domain is \ac{D-PSGD}~\cite{lian2017can}, in which all nodes exchange their models with a subset of neighbors every round.
\ac{DL} is increasingly being adopted in high-stakes domains such as healthcare and finance~\cite{warnat2021swarm}.

However, \ac{D-PSGD} and its derivatives are ill-suited for the \ac{OLTR} setting for three reasons.
First, they assume that each node's training data remains fixed during training, whereas in \ac{OLTR}, local datasets grow incrementally as users issue queries and generate clicks.
Second, \ac{D-PSGD} typically assumes a fixed communication topology, which is difficult to maintain in large decentralized networks where nodes join and leave freely and no central authority exists to coordinate topology management. 
Third, \ac{D-PSGD} requires all nodes to exchange models synchronously every round, which presupposes that nodes are continuously available.
In practice, users issue queries at irregular intervals and may leave and rejoin the network freely, a phenomenon known as \emph{churn}.

\textbf{\Acf{GL}}~\cite{ormandi2013gossip} is a class of \ac{DL} algorithms in which nodes independently and periodically exchange their model with randomly selected nodes in the network.
This process relies on a \ac{RPS} to sample communication partners.
Unlike \ac{D-PSGD}, \ac{GL} is fully asynchronous: nodes contribute to and benefit from the federation whenever they are active, without waiting for other nodes to synchronize their models with.
This makes \ac{GL} naturally churn-resilient, as nodes joining or leaving the network do not stall the training process.
Communication in \ac{GL} is push-based: upon completing a local update, a node pushes its current model to a small set of sampled nodes, which aggregate the received model into their own upon receipt.
These properties make \ac{GL} a natural fit for the \ac{OLTR} setting.


\subsection{Poisoning Attacks in Collaborative Learning}
A key concern in any collaborative \ac{ML} system are poisoning attacks: nodes can send maliciously crafted updates to degrade the model utility of other nodes.
In the context of \ac{OLTR}, such attacks can degrade ranking quality across the network, harming user experience or steering users toward irrelevant or manipulated results.
Poisoning attacks have been studied extensively in \ac{FL}, where a central server observes all client updates simultaneously and can apply population-level statistical filters to detect and discard outliers~\cite{fang2020local}.
In \ac{DL}, however, such server is not available and nodes instead receive updates only from a small, changing subset of other nodes and must assess their trustworthiness independently.
While some robustness algorithms have been proposed for \ac{DL}~\cite{he2022byzantine,gaucher2024unified,fang2024byzantine}, they are designed for synchronous \ac{D-PSGD}-like protocols.
Robustness against poisoning in \ac{GL}-like algorithms, where communication is asynchronous, push-based, and driven by irregular user activity, remains an open problem.
Thus, the research question that this work answers is: \emph{How can a node in a decentralized, gossip-based \ac{OLTR} system reliably distinguish between beneficial and malicious model updates, without access to a trusted server, or knowledge of which peers are malicious?}





%% file: sections/doltr.tex
\section{The Design of \sys}
\label{sec:rankguard}

\begin{figure*}[t]
    \centering
    \includegraphics[width=.92\linewidth]{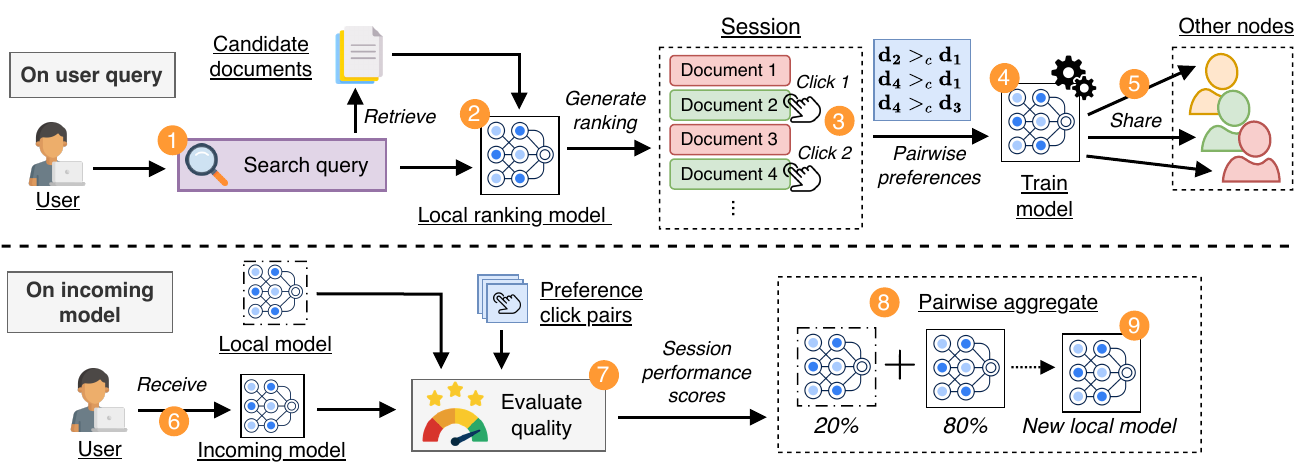}
    \caption{The workflow of \sys, our decentralized framework for robust \ac{OLTR}.}
    \label{fig:workflow}
\end{figure*}

We present \sys, a novel and decentralized mechanism for \ac{OLTR} that provides robustness against malicious nodes that attempt to corrupt the ranking of other nodes by spreading malicious model updates.
The \sys protocol is shown in \Cref{alg:rankguard} and the full workflow is visualized in \Cref{fig:workflow}.

\subsection{System and Threat Model}
Before describing the design of \sys, we outline our system and threat model.

\textbf{Network and nodes.}
We consider a network of $n$ nodes, each interacting with a search engine through queries and clicks.
Nodes are connected through a dynamic communication graph in which every node knows about $\lceil \log_2 n \rceil$ other nodes.
This is a commonly chosen fanout value in the domain of \ac{DL} that is known to converge well, and therefore follows standard practice~\cite{de2023epidemic}. 
Each node has access to a \ac{RPS} which is a service that provides nodes with a renewed set of nodes, sampled uniformly from the network population~\cite{guerraoui2024peerswap}.
We assume the \ac{RPS} delivers an unbiased random sample of peers and that messages between honest users are delivered intact.
Recent work shows that continuously refreshing the set of neighboring nodes shrinks the attack surface of malicious nodes~\cite{belal2025inferring, touat2025exposing}.
Importantly, our work does not require any central server and all components of \sys and dependencies can be implemented using decentralized primitives.

\textbf{Search sessions.}
When a user issues a query $q$, they are presented with a ranked list of candidate documents $R$.
The user examines $R$ from top to bottom and clicks documents according to a click model.
The click model determines, based on each document's true relevance label, the \emph{click} probability and a \emph{stop} probability (\ie, the chance that the user ends the session after a click and examines no further documents).
Click models~\cite{chapelle2009dynamic} are a standard tool for reproducing realistic behavior patterns such as position bias and noisy, relevance-dependent clicks without requiring live users.
We simulate search sessions in round-robin order across all nodes; in each round, every node processes a single session.

\begin{algorithm}[t]
\input{algorithms/doltr}
\caption{\sys from the perspective of node $u$}
\label{alg:rankguard}
\end{algorithm}

\textbf{Threat model.}
We assume that there is a fraction of malicious nodes $\beta$ in the network: they deviate from the protocol in \Cref{alg:rankguard} by sending poisoned model updates.
We make no assumption about which nodes are malicious, and honest users do not know $\beta$. 
For evaluation purposes, however, baselines that do require $\beta$ are given its true value, granting them an advantage that our system does not require.
The set of malicious nodes remains fixed for the duration of each experiment run.

We focus on \emph{untargeted} poisoning attacks, in which malicious nodes aim to indiscriminately degrade ranking quality across all honest users.
Untargeted poisoning attacks constitute a natural worst-case for an availability-oriented threat model: a successful attacker prevents the federation from converging to a useful ranking model, nullifying the benefits of collaboration.
More specifically, malicious nodes follow the protocol but send carefully crafted model updates to other nodes (Step~{5} in \Cref{fig:workflow}).

We acknowledge that the threats in decentralized systems extend beyond poisoning attacks, \eg, Sybil attacks against the \ac{RPS}~\cite{douceur2002sybil} or eclipse attacks~\cite{singh2006eclipse}.
Deploying defenses on the network layer is an orthogonal problem that can be solved by using robust peer samplers~\cite{auvolat2023basalt,belal2025granite}.
Similarly, we consider privacy attacks (\eg, membership inference~\cite{touat2025exposing}) on the datasets of individual nodes beyond our scope.
Our threat model therefore concerns the effectiveness of the trained ranking model, not the confidentiality of training data.

\subsection{\sys in a Nutshell}
\label{subsec:nutshell}

The key question when doing \ac{OLTR} using \ac{GL} is how a node should weigh a single model received from another node when there is no trusted server that has access to multiple models to run and no way to tell which nodes are malicious.
\sys answers this by treating each node's own click history as a private, trusted reference.
Our main insight is that an honest node accumulates reliable records of which documents they clicked on past result pages.
While this record may be noisy (\eg, because a user misclicks or misinterprets search results), it cannot be exploited by an attacker.
So instead of asking whether a received model resembles the models sent by other nodes, as done by traditional statistical defenses~\cite{blanchard2017machine,yin2018byzantine,he2022byzantine}, a node using \sys determines \emph{whether the received model explains its own past clicks better than its current model does.}
A model that genuinely improves the local ranking quality is aggregated into the local model, whereas one trained to disrupt honest rankings is effectively discarded.
\sys does so by efficiently computing session performance scores, after correcting for position bias.
This pairwise model judgment is made entirely using the user's local click history. \sys therefore needs neither separate validation data nor any knowledge of how many peers are malicious.
It also avoids a common failure mode of statistical defenses, where honest updates can be diverse enough to be mistaken for malicious ones, particularly when local datasets are very diverse between nodes. \mg{cite would be nice}

\subsection{The \sys Workflow}
\label{subsec:workflow}


We next present the full workflow of \sys, primarily driven by two events: a node performing a search (and clicking items), and a node receiving a model from another node.
Importantly, every node $u$ maintains a local ranking model $\theta_u$  trained on their own click stream via \ac{PDGD} (see \Cref{sec:pdgd}).
$\theta_u$ is randomly initialized by node $u$ and is locally optimized with their session history $Q_u$. Each $q \in Q_u$ stores the candidate documents $D_u^{(q)}$, click pair preferences $\mathcal{P}_u^{(q)}$, and for each $\bm{d_i} >_c \bm{d_j} \in \mathcal{P}_u^{(q)}$, the click pair bias $\rho_{ij}^{(q)}$. 

\subsubsection{On user search.}
When node $u$ performs a search query $q$ (Step 1 in \Cref{fig:workflow}), the search engine first retrieves a candidate document set $D_u^{(q)}$ that is relevant to the search query (line 4 in \Cref{alg:rankguard}).
We remark that these documents may not be stored locally by node $u$ and could be fetched by other nodes, \eg when using decentralized file-sharing applications such as \textsc{IPFS}~\cite{benet2014ipfs}.
Documents in $D_u^{(q)}$ are then ranked by $\theta_{u}$, producing a ranked list $R^{(q)}$ (Step~{2}).
Node $u$ then examines the list and clicks on one or more documents (Step~{3}).
When the search session is completed, the node first infers a set of pairwise preferences $\mathcal{P}_u^{(q)}$ while also estimating the position bias $\rho_{ij}^{(q)}$ for each click pair, then trains $\theta_{u}$ locally on these pairwise preferences, and updates the resulting model using \ac{PDGD} with learning rate $\eta$ (Step~{4}).
After this, node $u$ samples $\lceil \log_2 n \rceil$ other nodes from the network, using the \ac{RPS} service, and sends the updated model $\theta_u$ to these sampled nodes.



\subsubsection{Receiving a model.}
When node $u$ receives a model from another node $v$ (Step {6} in \Cref{fig:workflow}), it will judge the quality of the received model and aggregate it if the received model has good quality.
We do so by replaying both the local model $\theta_u$ the received model $\theta_{v}$ on the node's click history to assess whether, and by how much, $\theta_{v}$ outperforms the user's current model $\theta_{u}$ (Step {7}).
Note that past clicks do not constitute clean relevance signals, as they are influenced by position bias.
Furthermore, recall that rankings are generated from a probabilistic distribution.
Thus, the produced rankings by the evaluator are partly shaped by chance.
A good evaluation function should therefore answer:
\emph{
    What is the probability that the received model produces a ranking consistent with the user's preferences, after correcting for position bias?}

To explain the design of our evaluation function, we first assume that click signals are unbiased.
For each $u\in[n]$, node $u$'s model performance in a single search session $q\in Q_u$ can then be measured as the probability that all clicked documents are sampled before the examined non-clicked documents, and therefore ranked above them.
For all $u\in[n]$, we can express this as the sum of log-probabilities:
\begin{equation}\label{eq:simpleQueryScore}
   P(q\mid\theta_u, q\in Q_u)=\sum_{\bm{d_i}>_c \bm{d_j}\in\mathcal{P}_u^{(q)}}\log P(\bm{d_i}\succ \bm{d_j}\mid D_u^{(q)},\theta_u)
\end{equation}
In other words, $P(q\mid\theta_i,q\in Q_u)$ quantifies how well the local model of each node explains a given search session.
However, the click preferences inferred from the search session are biased.
From \Cref{eq:rho}, we know how to quantify this bias.
The bias is neutral (\ie the preference pair $\bm{d_i}>_c \bm{d_j}$ is \emph{unbiased}) when $\rho^{(q)}_{ij}=0.5$.
A value close to $0$ indicates that the position bias favored $\bm{d_i}$, whereas a value close to $1$ indicates that it favored $\bm{d_j}$.
A value close to $1$ yields a stronger relevance signal: the node preferred $\bm{d_i}$ despite the bias favoring $\bm{d_j}$.

We propose to weigh each click preference pair by $\rho^{(q)}_{ij}$.
This, for any node $u$, gives us the \emph{session performance score} $S(q, \theta_u)$ for search session $q$ using model $\theta_u$:
\begin{equation}
    S(q, \theta_u)=\sum_{\bm{d_i}>_c \bm{d_j}\in\mathcal{P}_u^{(q)}}\rho^{(q)}_{ij}\log P(\bm{d_i} \succ \bm{d_j}\mid D_u^{(q)},\theta_u)
    \label{eq:session_score}
\end{equation}
Thus, a high predicted probability for $\bm{d_i}\succ \bm{d_j}$ contributes less to the score when $\bm{d_i}$ already had a positional advantage.
If a received model $\theta_v$ from node $v$ has a better average session performance score than the local model $\theta_{u}$ of node $u$, we assume it better captures the node's preferences.
We quantify this improvement as 
\begin{equation}
    \Delta S =
    \frac{1}{|Q_u|}
    \sum_{q\in Q_u}
    \left[
        S(q,\theta_{j})
        -
        S(q,\theta_{i})
    \right].
\end{equation}
A positive value of $\Delta S$ indicates that the received model better explains the node's click preferences, corrected for position bias.
However, we argue that $\Delta S$ should not be treated as a binary accept or reject criterion.
Even a negative value of $\Delta S$ may carry useful signal if the query history is small and the assessment is therefore uncertain.
To quantify the confidence of our performance evaluation, we compute a per-query score
$
X_q=S(q, \theta_v)-S(q, \theta_u).
$
We then derive the one-sample $t$-statistic $t=\sqrt{|Q_u|}\bar{X}/s_X$, which accounts for both the magnitude of the improvement and the number of queries.
\TODO{introduce t-plot}
We then map $t$ to an interpolation weight $\alpha=\sigma(\kappa t)$, where $\sigma$ is the sigmoid function and $\kappa$ controls sensitivity, and update the local model $\theta_{u}$ as
$
   \theta_{u}^{+} \leftarrow (1-\alpha)\cdot\theta_{u} + \alpha\cdot\theta_{v}.
$
Thus, we aggregate the incoming model into the local model, weighted by the performance of the incoming model (Steps 8 and 9).
We find that $\kappa=4$ works best across datasets, click models, and attacks.
When the value of $\Delta S$ is large and well-supported by many queries, the value of $t$ is large, $\alpha$ approaches 1, and $\theta_{v}$ largely replaces $\theta_{u}$.
When evidence is weak or contradictory, $\alpha$ remains near $0$ and the local model is mostly preserved.
In this way, \sys effectively turns each node's own interaction history into a shield: malicious models, trained against the node's interests, will score poorly on that history and be absorbed with negligible weight.

%% file: algorithms/doltr.tex
\DontPrintSemicolon

\newcommand{\Call}[2]{\FuncSty{#1}(#2)}
 
\SetKwProg{Init}{Initialize}{}{}
\SetKwProg{OnQuery}{OnQuery}{}{}
\SetKwProg{OnReceive}{OnReceive}{}{}

\Init{}{

    $\theta_u \leftarrow$ \Call{RandomInit}{} \\
}

\OnQuery{query $q$}{
    $D_u^{(q)} \gets $ \Call{RetrieveDocuments}{$q$} \\
    $R^{(q)} \gets $ \Call{RankDocuments}{$\theta_u, D^{(q)}$} \\
    $\mathcal{P}_u^{(q)} \gets \Call{GeneratePairwiseClickPreferences}{R^{(q)}}$ \\
    
    $\nabla f_{\theta_u} \leftarrow \bm{0}$ \\
    \ForEach(){$\bm{d_i} >_c \bm{d_j} \in \mathcal{P}^{(q)}$}{
        $\rho_{ij}^{(q)} \gets \Call{EstimateBias}{\bm{d_i},\bm{d_j},R,D^{(q)}}$ \tcp*{\Cref{eq:rho}}
        $\nabla f_{\theta_u} \gets \nabla f_{\theta_u} + \rho_{ij}^{(q)} \nabla P(\bm{d_i} \succ \bm{d_j} \vert D^{(q)},\theta_u)$ \\
    }
    $\theta^{+}_{u} \leftarrow \theta_{u} + \eta \nabla f_{\theta_{u}}$ \\
    \ForEach{$p \in$ \Call{SamplePeers}{}}{
        \Call{Send}{$p,\ \theta_{u}$}\;
    }
    $Q_u \gets Q_u \cup \{q\}$ \\
}
 
\OnReceive{model weights $\theta_{v}$}{
    \For{$q \in Q_u$}{
    
        $X_q \gets S(q,\theta_v)-S(q,\theta_u)$ \tcp*{\Cref{eq:session_score}}
    }
    $t \gets \sqrt{|Q_u|} \cdot \bar{X} / s_X$\;
    $\alpha \gets \sigma(\kappa t)$\;
    $\theta^{+}_{u} \leftarrow (1-\alpha) \cdot \theta_{u} + \alpha \cdot \theta_{v}$ \\
}

%% file: sections/theory.tex
\section{Convergence Analysis of \sys}\label{sec:theory}

One of the key non-trivialities in deriving the convergence bound for \sys lies in the fact that it cannot simply adapt the analysis of existing \ac{DL} algorithms. Because each node $u$ independently computes a dynamic weight for incoming models in every local round $t$ using a statistical $t$-test, the underlying aggregation scheme is fundamentally altered, breaking standard machinery used in typical \ac{DL} convergence analysis, and, in turn, requiring a novel approach to handle the shifting properties of the mixing matrix.
We now theoretically analyze the convergence of \sys. 
In the interest of space, we present the sketch proof of \Cref{th:convergence}, deriving the convergence rate of \sys, in Appendix~\ref{app:convergence_proof}. 
We proceed by making the following standard assumptions on the learning objective, in line with the existing literature of convergence analysis in \ac{DL}~\cite{koloskova2020unified,de2023epidemic,even2024asynchronous,biswas2025noiseless,biswas2025boosting,biswas2026your}.

\begin{assumption}[Smoothness]\label{assump:smoothness}
For any model $\theta$, the scoring function $f_{\theta}$ is differentiable and
there exists $L < \infty$ s.t. for all $x, y \in \mathbb{R}^d$, $ ||\nabla f_{\theta}(y) - \nabla f_{\theta}(x)|| \le L ||y - x||$.
\end{assumption}

\begin{assumption}[Bounded noise]\label{assump:bounded_noise}
For any model $\theta$, there exists $\sigma < \infty$ s.t. for every node $u$, we have: 
\begin{equation}
   \mathbb{E}_{x\in {D}_u^{(q)}} \left[ ||\nabla f_{\theta}(x) - f_{\theta}(x)||^2 \right] \le \sigma^2.\nonumber
\end{equation}
\end{assumption}

\begin{assumption}[Bounded heterogeneity]\label{assump:bounded_heter}
For any model $\theta$, there exists $\mathcal{H} < \infty$ such that for all $x \in \mathbb{R}^d$, we have:
\begin{gather*}
\frac{1}{n}\sum_{u\in[n]}
\mathbb{E}_{x\sim D_u^{(q)}}
\left[
\left\|
\nabla_{\theta} f_{\theta}(x)
-
\nabla_{\theta} F_q(\theta)
\right\|^2
\right]
\le \mathcal{H}^2,
\\
\text{where}\quad
F_q(\theta)
=
\frac{1}{n}\sum_{u\in[n]}
\mathbb{E}_{x\sim D_u^{(q)}}
\bigl[f_{\theta}(x)\bigr].
\end{gather*}
\end{assumption}
In addition to Assumptions~\ref{assump:smoothness}, \ref{assump:bounded_noise}, and~\ref{assump:bounded_heter}, we assume that, for every node, each session comprises exactly one click. Under these above assumptions, we formalize the workflow of \sys and define the notion of a \emph{local round} for each node as follows. Let the local model of any node $u\in [n]$ in its local round $t_u\geq 0$ be denoted by $\theta_u^{(t_u)}$. Then, node $u$ in its local round $t_u$ $(a)$ receives $\theta_v^{(t_v)}$ from another node $v$ in the network operating in its local round $t_v$ (can be different from $t_u$), $(b)$ performs the weighted aggregation to incorporate $\theta_v^{(t_v)}$ into its own local model $\theta_u^{(t_u)}$ to produce $\theta_u^{(t_u+\frac{1}{2})}=(1-\alpha_{uv}^{(t_u)})\cdot\theta_u^{(t_u)} + \alpha_{uv}^{(t_u)}\cdot\theta_v^{(t_v)}$, where $\alpha_{uv}^{(t_u)}$ is computed as described in line 19 in \Cref{alg:rankguard}, and $(c)$ finally performs the local optimization using \ac{PDGD} to obtain $\theta_u^{(t_u+1)}=\theta_{u}^{(t_u+\frac{1}{2})}+\eta\nabla f_{\theta_u^{(t_u+\frac{1}{2})}}$ as given in lines 3-12 in \Cref{alg:rankguard}.  We are now in a position to state the main convergence result of \sys.

\begin{theorem}[Convergence of \sys]\label{th:convergence}
    Let $\Delta_0\in\mathbb{R}$ s.t. $\Delta_0\geq\max_{u\in[n]}\max_{\bm{d}\in D_u^{(q)}}\left\{F_{\theta_u^{(0)}}(\bm{d})-\min_{\theta}{F_{\theta}}(\bm{d})\right\}$. If all the above assumptions hold and if for all $u,v\in [n]$ and $t_u>0$, $\mathbb{E}\left[\alpha_{uv}^{(t_u)}\right]=\hat{\alpha}$ and $\mathrm{Var}\left[\alpha_{uv}^{(t_u)}\right] = \sigma_{\hat{\alpha}}^2$, then, for a constant step-size $\eta$, we have
 \begin{align}
    &\frac{1}{n}\sum_{u\in[n]}\frac{1}{T}\sum_{t_1,\ldots,t_n=0}^{T-1}\mathbb{E}_{\bm{d}\in D_u^{(q)}}\left[\left\lvert\left\lvert\nabla F_{\theta_u^{(t_u)}}(\bm{d})\right\rvert\right\rvert^2\right]\in\nonumber\\
    &\mathcal{O}\left(\sqrt{\frac{L\Delta_{0}}{nT}}+ \left(\frac{L^2\Delta^2_{0}}{s^*T^2}\left(1-\frac{s^*}{n-1}\right)^n-\frac{L^2\Delta^2_{0}}{(n-1)T^2}\right)^{1/3}
    +\frac{L\Delta_{0}}{T}\right),\nonumber\\
    &\text{where}\quad s^* = \frac{2\hat{\alpha}(1 - \hat{\alpha}) - 2\sigma_\alpha^2}{(1 - \hat{\alpha})^2 + \hat{\alpha}^2 + 2\sigma_{\hat{\alpha}}^2}.\nonumber
\end{align}
\end{theorem}

\begin{remark}\label{rem:interpret_convergence}
    The convergence rate derived in \Cref{th:convergence} essentially demonstrates how the different design aspects of \sys fundamentally interact with the theoretical convergence guarantees of \sys. Crucially, we note that having $\hat{\alpha}=1/2$, \ie if the nodes aggregate their own models and the received models with an equal weight without performing any \emph{quality check} of the received models, we obtain $s^*=1$, thus boiling down to the exact convergence rate of \elloc. 
    Conversely, if $\hat{\alpha}$ is $0$ or $1$, \ie the nodes \emph{never} incorporate the models they receive, or \emph{always} replace their local models with the received models, respectively, the convergence rate, in expectation, blows up, ruining the utility, as we would expect. Moreover, we observe that, keeping everything else constant, the convergence rate's asymptotic dependence on aggregation weight determined by each node to validate the quality of the received models is via $\mathcal{O}\left(\exp{-\frac{2\hat{\alpha}(1 - \hat{\alpha}) - 2\sigma_\alpha^2}{(1 - \hat{\alpha})^2 + \hat{\alpha}^2 + 2\sigma_{\hat{\alpha}}^2}}\right)$ as the total number of nodes in the network tends to infinity. 
    Finally, it is noteworthy that the convergence rate of \sys is directly proportional to the smoothness factor $L$ of each node's local objective function, and the maximum initialization error $\Delta_0$, as is observed in convergence rates of standard \ac{DL} algorithms. 
\end{remark}







%% file: sections/experimental_setup.tex
\section{Experimental Setup}
\label{sec:exp_setup}

We evaluate the robustness and efficiency of \sys against four poisoning attacks and compare with five state-of-the-art baselines.
This section describes our experimental setup, including attacks, defense baselines, datasets, and implementation details.
Throughout this section, we use $u$ to refer to the node receiving a model update, and $v$ to refer to the node sending it to $u$.


\subsection{Poisoning Attacks}
\label{sec:attacks}

We run our experiments using one data poisoning attack native to \ac{OLTR} and two state-of-the-art model poisoning attacks that have become standard benchmarks in the Byzantine-robust learning literature~\cite{fang2020local,belal2025granite,yang2024byzantine,he2022byzantine}.
We also experiment with a stronger \emph{adaptive attack}.
Together, these attacks span the spectrum from weak, realistic threats to strong, oracle-knowledge attacks.

\textbf{Data Poisoning (\flip).}
The adversarial click model introduced in \citet{wang2023analysis} is an attack in federated \ac{OLTR}.
Malicious nodes simulate interactions using a \emph{poison} click model in which click probability is inversely proportional to document relevance: highly irrelevant documents are clicked with near certainty, while relevant documents are almost never clicked. 
The resulting pairwise preferences push \ac{PDGD} updates away from the honest gradient; the corrupted model weights are then pushed to other nodes.
This attack assumes no knowledge by the adversary about the victim node, making it a realistic lower bound on attacker capability.

\textbf{Little Is Enough (\lie).}
The \lie attack exploits the coordinate-wise variance among honest peer models to craft perturbations that remain statistically indistinguishable from benign contributions~\cite{baruch2019little}.
The main idea is that statistics-based defenses assume malicious updates will be outliers; \lie invalidates this assumption by operating \emph{within} the empirical spread of honest updates, bounding the perturbation to stay within the defenders' acceptance region while maximally shifting the aggregate in a chosen direction.
Concretely, attackers compute the coordinate-wise mean $\mu_j$ and standard deviation $\sigma_j$ of the honest peers' current models and transmit $ (\theta_\text{mal})_j = \mu_j - z^{\max} \cdot \sigma_j, $ where $z^{\max}$ is a perturbation factor derived from $n$ and $\beta$ such that a majority of honest models still lies farther from $\mu$ than $\theta_\text{mal}$, ensuring defenses preferentially retain the poisoned values.
Because \sys is designed for pairwise aggregation, there is no population of incoming models from which $\mu$ and $\sigma$ could be derived.
Instead, we let the attacker derive those statistics from the current models of all honest nodes in the network.
This has important implications about the quality of the attack.
Recall that in each round, every node processes a single session in round-robin order.
In expectation, therefore, all honest nodes start a round with equally good models; by the end of a round, most honest nodes are only one session ahead of the current node.
Because honest nodes are near-synchronized, the LIE perturbation reduces to a near-honest model plus a small noise term, rendering the attack closer to Gaussian noise.

\textbf{Inner Product Manipulation (\ipm).}
Similar to \lie, \ipm~\cite{xie2020fall} directly targets Byzantine-robust aggregation methods.
The \ipm attack evades statistics-based defenses by sending model updates of bounded magnitude that point opposite to the honest gradient direction.
When such updates are averaged with honest contributions, the aggregated update can have a negative inner product with the true gradient, causing the model to ascend rather than descend the loss.
\citet{xie2020fall} show that this is especially damaging late in training: as the honest gradient shrinks near convergence, even small adversarial deviations suffice to flip the sign of the aggregate.
We adapt \ipm to our setting and assume a malicious node $v$ knows the local model $\theta_{u}$ of victim node $u$.
Node $v$ trains $\theta_v$ on a single search session to obtain a gradient $g$, and updates $\theta_{v}$ using $-\epsilon g$ instead of $g$.
Node $v$ then sends $\theta_{v}$ to victim $u$.
The victim thus receives an update that opposes the direction of honest learning.
We use $\epsilon = 10$, as preliminary experiments showed only marginal ranking degradation at smaller values.

\textbf{Adaptive Attack (\adapt).}
To stress-test the robustness of \sys, we develop an \emph{adaptive attack} inspired by \citet{fang2020local}, who formulate model poisoning as an optimization problem against a known defense.
Our goal is to establish an upper bound on attacker capability under full knowledge: the attacker has access to the victim's model $\theta_u$ and session history $Q_u$.

We let the attacker craft a poisoned model by minimizing a loss with two terms.
The first term rewards alignment of the induced update $\theta_v-\theta_u$ with a harmful direction $\hat{g}$, which we define as the negative of the victim's honest descent direction.
The attacker computes this exactly by simulating the victim's next PDGD step on $Q_u$, and taking its opposite, yielding the harmful gradient:
$$
\hat{g} = -\nabla_\theta \sum_{q \in Q_u} S(q, \theta) \,\Big|_{\theta = \theta_u}
$$
We measure alignment as the negative cosine similarity and define the loss term 
$$
\mathcal{L}_\text{harm}(\theta_v) = -\frac{\langle \theta_v - \theta_u,\, \hat{g}\rangle}{\|\theta_v - \theta_u\|\,\|\hat{g}\| + \epsilon},$$
where $\epsilon > 0$ is a small constant to prevent division by zero. 
A second loss term penalizes rejection by \sys:
$$
\mathcal{L}_\text{accept}(\theta_v) = \big(1 - \alpha(\theta_v)\big)^2
$$
For this, let $\alpha(\theta_v)$ denote the weight $\alpha$ the victim assigns to $\theta_v$ given $\theta_u$ and $Q$.
Finally, the attacker solves
$$
    \theta_v^* = \arg\min_{\theta_v} \; \mathcal{L}_\text{harm}(\theta_v) + \mathcal{L}_\text{accept}(\theta_v)
$$

Practically, the attack proceeds in two phases:
\begin{enumerate*}
    \item We first warm-start $\theta_v$ by overfitting on $Q$ from $\theta_u$ until the loss stabilizes.
At this point, $\theta_v$ already ranks $Q$ well, so the optimizer can move it in harmful directions without immediately collapsing~$\alpha$. 
    \item We then optimize $\theta_v^*$. 
\end{enumerate*}
Both phases use the same gradient update rule and learning rate that is used for honest PDGD training.
Our implementation runs $100/|Q|$ epochs for the warm-start phase and 50 steps for the loss optimization; preliminary experiments showed both losses stabilized within these budgets.
The resulting $\theta_v^*$ is sent to the victim, concluding one round of the adaptive attack.
Repeated across multiple rounds, this procedure constitutes the strongest poisoning attack against \sys that we consider.

\subsection{Defense Baselines}
\label{sec:defenses}
We compare \sys against five defense baselines: three defenses based on robust statistics and two defenses based on a private reference dataset.
Together, these five baselines span the two dominant paradigms in Byzantine-robust aggregation: statistical filtering and trusted-reference scoring.
Thus, they comprise a representative point of comparison against \sys.

\textbf{CS, GTS, and CWTM.}
We include three state-of-the-art methods based on robust statistics: CS, GTS, and CWTM.
CS clips each update toward a robust center before averaging; GTS discards the updates farthest from that center; CWTM trims the top and bottom $\beta$ fraction per coordinate and averages the rest.
Note that these algorithms require a pool of candidate models to make aggregation decisions, typically assuming knowledge of the fraction of malicious models $\beta$.
To accommodate them, we let nodes buffer incoming models and, once the buffer reaches size $\lceil\log_2 n\rceil$, apply the respective aggregation rule and clear the buffer.
We assume that $\beta$ is known by all nodes, so that their performance demonstrates an upper bound on the defenses' capability.

\textbf{\fltrust.}
When a node receives a model $\theta_v$, it computes a candidate update $g = \theta_v - \theta_u$.
It then crafts a trusted reference update $r = \hat{\theta}_u - \theta_u$ where
$\hat{\theta}_u$ is created by training $\theta_u$ for one epoch on the local session history $Q_u$.
Subsequently, it computes the ReLU-clipped cosine trust score
$
\tau = \max\left(0, \frac{\langle g, r \rangle}{\|g\|\|r\|}\right).
$
If either update has zero norm or $\tau=0$, candidate $g$ is rejected.
Otherwise, $g$ is rescaled to match the magnitude of the trusted reference and applied:
$\theta_u \gets \theta_u + g\cdot\|r\|/\|g\|$.
In the original \fltrust protocol, trust scores weight a server-side aggregate over multiple client updates; in \sys, where received models are processed pairwise, we use the score as an acceptance criterion.

\textbf{\zenops.}
We adapt the score-based validation rule from ZenoPS~\cite{xie2022zenops}.
Using the same candidate update $g$ and trusted reference update $r$ as used by \fltrust, the candidate is \emph{rejected} if its alignment with the reference is too small, $\langle g, r \rangle < \zeta
\|r\|^2 + e$.
Otherwise, the update $g$ is clipped to $\|\tilde{g}\|^2 \le (1+\gamma)\|r\|^2$ and applied: $\theta_u \gets \theta_u + \tilde{g}$.
In line with the original paper, we set $\zeta=-0.02$, $\gamma=0$, and $e=0$.




\begin{table*}[t]
    \centering
    \caption{Robustness after 30\,000 sessions in a 100-node network with 20 malicious nodes.
    The best and second-best baselines in each setting are highlighted in bold and underlined, respectively.}
    \input{tabs/results}
    \label{tab:robustness}
\end{table*}

\subsection{Datasets and Click Models}
We evaluate \sys and baselines using four datasets and three click models.
Due to space constraints, some experiments report only a representative setting: WEB30K with the \emph{perfect} click model, and an attack share  $\beta=0.2$.

\textbf{Datasets.}
We adopt four standard datasets commonly used in \ac{LTR}: WEB30K, MQ2007~\cite{qin2013introducing}, Yahoo~\cite{chapelle2011yahoo}, and Istella~\cite{dato2022istella22}.
MQ2007 is the smallest with coarse three-level relevance labels, while WEB30K, Yahoo, and Istella offer larger query sets with five-level relevance. Yahoo provides the highest-dimensional feature space ($d=700$ features), and Istella the largest candidate sets per query.

\textbf{Click models.}
Click models provide a probabilistic account of how users interact with a ranked list, and are widely used to simulate user behavior in \ac{OLTR} settings.
Under the \ac{SDBN} model~\cite{chapelle2009dynamic}, a user examines documents from top to bottom, clicking each with a probability conditioned on its relevance $\mathrm{rel}(d)$, and stopping after a click with a relevance-dependent stopping probability.
In this work, we consider three instantiations of \ac{SDBN}, describing different user behaviors: a \emph{perfect} user who clicks all relevant documents and rarely stops, a \emph{navigational} user seeking a single relevant result, and an \emph{informational} user who clicks broadly and provides the noisiest signal.
These click models are commonly considered in this domain~\cite{wang2023analysis}.

\subsection{Implementation and Evaluation Metrics}
We simulate a network of $n=100$ nodes that follow the protocol described in \Cref{alg:rankguard}.
The parameters of each node's local model are initialized randomly.
In each iteration, users take turns in round-robin order, each sampling a query from the dataset, ranking candidate documents using their model, and generating clicks via a click model.
The configuration of the click models follows standard practice (for details, see Table 1 in~\cite{wang2023analysis}).
Based on the clicks, the respective node updates their model and sends it to $\lceil \log_2 n \rceil=7$ randomly sampled nodes, using a \ac{RPS}.
For local model training, we employ a learning rate of $\eta=0.1$.
By default, we assume a linear model $f_\theta(\bm{d})= \theta^\top \bm{d}$, where the model size equals the feature dimensionality of the respective dataset, so $\theta\in\mathbb{R}^d$.
To verify that \sys is not tied to this choice, we additionally benchmark a non-linear (\emph{neural}) ranker: a feed-forward network with a single hidden layer of 64 units and sigmoid activation.
The architecture and hyperparameter choices are adopted from prior work~\cite{oosterhuis2018differentiable,wang2021effective}.
Unless otherwise stated, we assume that 20\,\% of nodes in the network are malicious (\ie, $\beta=0.2$).

\textbf{Implementation and hardware.}
All baselines and experiments are implemented in Python.
For \ac{PDGD}, we adopt the \textsc{NumPy}-based implementation from \citet{wang2021effective}.
Our experiments, including network simulations, are run on a single machine with an AMD EPYC 7282 processor; no GPU required.
The efficiency benchmark (\Cref{sec:exp_efficiency}) uses a single CPU core, with a warm-up period that is excluded from the measurements.
We always report mean and standard deviation across repetitions: across nodes for network simulations, and across random seeds (count stated per experiment) for standalone benchmarks.
To support reproducibility, we make our code publicly available.\footnote{See \url{https://anonymous.4open.science/r/rankguard-B569}.}

\textbf{Ranking accuracy.}
Ranking quality is typically evaluated offline using Normalized Discounted Cumulative Gain (nDCG)~\cite{jarvelin2002cumulated}. 
nDCG measures how well a ranked list matches an ideal ordering by grading each document by its relevance label and discounting contributions by position $\text{nDCG@}k = \frac{1}{Z_k} \sum_{i=1}^{k} \frac{2^{\mathrm{rel}(\bm{d_i})} - 1}{\log_2(i+1)}$,
where $\mathrm{rel}(\bm{d_i})$ is the true relevance label of $\bm{d_i}$ and $Z_k$ is a normalisation constant such that a perfect ranking scores 1.
Following common practice, we truncate nDCG at $k=10$, as users rarely examine results beyond the first page. 
We evaluate each honest user's model on a held-out test set
after every full round (100 sessions), reporting the mean and standard deviation across nodes.

%% file: tabs/results.tex
\resizebox{\textwidth}{!}{%
\setlength{\tabcolsep}{2.8pt}
\begin{tabular}{c|lcccccccccccc}
\toprule
\multicolumn{1}{c}{} & Defense & \multicolumn{3}{c}{WEB30K} & \multicolumn{3}{c}{MQ2007} & \multicolumn{3}{c}{Yahoo} & \multicolumn{3}{c}{Istella} \\
\cmidrule(lr){3-5} \cmidrule(lr){6-8} \cmidrule(lr){9-11} \cmidrule(lr){12-14}
\multicolumn{1}{c}{} & & perfect & navigat. & informat. & perfect & navigat. & informat. & perfect & navigat. & informat. & perfect & navigat. & informat. \\
\midrule
\rowcolor{gray!20} \cellcolor{white} & \textit{Oracle} & \textit{0.415(2)} & \textit{0.375(3)} & \textit{0.335(1)} & \textit{0.464(2)} & \textit{0.452(1)} & \textit{0.441(3)} & \textit{0.712(1)} & \textit{0.706(1)} & \textit{0.699(1)} & \textit{0.565(2)} & \textit{0.538(2)} & \textit{0.531(2)} \\
\rowcolor{gray!20} \cellcolor{white} & Local & 0.326(16) & 0.305(11) & 0.278(38) & 0.434(11) & 0.405(23) & 0.364(39) & 0.678(8) & 0.647(33) & 0.615(49) & 0.500(14) & 0.420(50) & 0.287(129) \\
\midrule
\multirow{7}{*}{\rotatebox[origin=c]{90}{\flip}} & \sys & \textbf{0.404(14)} & \textbf{0.380(24)} & \textbf{0.324(41)} & \textbf{0.472(7)} & \textbf{0.456(16)} & 0.399(43) & 0.691(10) & 0.676(17) & 0.622(47) & \textbf{0.557(3)} & \underline{0.509(15)} & 0.303(95) \\
 & \fltrust & \underline{0.403(9)} & 0.300(26) & 0.278(34) & \underline{0.460(3)} & 0.421(18) & 0.378(39) & 0.679(9) & 0.648(36) & 0.617(49) & 0.550(9) & \textbf{0.509(58)} & 0.278(137) \\
 & \zenops & 0.374(24) & 0.307(21) & 0.264(49) & 0.455(5) & 0.422(17) & 0.388(45) & 0.678(9) & 0.655(27) & 0.610(48) & \underline{0.556(4)} & 0.451(31) & 0.284(138) \\
 & GTS & 0.347(3) & \underline{0.321(1)} & \underline{0.317(1)} & 0.445(2) & \underline{0.442(3)} & 0.416(6) & \underline{0.703(2)} & \textbf{0.686(3)} & \underline{0.674(5)} & 0.521(4) & 0.447(4) & \underline{0.419(5)} \\
 & CS & 0.366(5) & 0.319(2) & 0.316(1) & 0.446(2) & 0.431(4) & \underline{0.426(5)} & \textbf{0.703(2)} & 0.672(6) & \textbf{0.677(4)} & 0.530(5) & 0.433(5) & \textbf{0.429(6)} \\
 & CWTM & 0.333(2) & 0.291(2) & 0.311(2) & 0.452(3) & 0.418(6) & \textbf{0.428(6)} & 0.701(2) & \underline{0.682(3)} & 0.664(5) & 0.523(4) & 0.421(4) & 0.418(6) \\
\rowcolor{gray!20} \cellcolor{white} & None & 0.089(3) & 0.094(5) & 0.107(10) & 0.147(2) & 0.146(3) & 0.158(4) & 0.443(2) & 0.444(2) & 0.448(7) & 0.004(1) & 0.004(1) & 0.007(4) \\
\midrule
\multirow{7}{*}{\rotatebox[origin=c]{90}{\lie}} & \sys & \underline{0.391(17)} & \textbf{0.369(19)} & \textbf{0.308(45)} & \textbf{0.474(7)} & \textbf{0.454(15)} & \textbf{0.400(37)} & \textbf{0.689(9)} & \textbf{0.663(20)} & \textbf{0.613(50)} & \underline{0.551(3)} & \textbf{0.489(16)} & 0.274(77) \\
 & \fltrust & \textbf{0.400(7)} & \underline{0.307(15)} & 0.272(43) & 0.453(4) & \underline{0.428(16)} & 0.374(36) & 0.679(10) & \underline{0.658(23)} & 0.608(51) & 0.546(12) & 0.440(28) & \underline{0.289(130)} \\
 & \zenops & 0.379(25) & 0.305(14) & \underline{0.273(44)} & \underline{0.460(7)} & 0.427(13) & \underline{0.376(39)} & \underline{0.680(8)} & 0.656(24) & \underline{0.613(53)} & \textbf{0.554(4)} & \underline{0.449(34)} & \textbf{0.291(134)} \\
 & GTS & 0.128(0) & 0.124(0) & 0.123(0) & 0.310(1) & 0.225(2) & 0.192(1) & 0.485(0) & 0.467(0) & 0.465(0) & 0.379(5) & 0.022(0) & 0.020(0) \\
 & CS & 0.247(3) & 0.144(3) & 0.124(0) & 0.354(2) & 0.272(3) & 0.197(1) & 0.595(3) & 0.472(0) & 0.467(0) & 0.494(5) & 0.336(6) & 0.023(0) \\
 & CWTM & 0.192(5) & 0.125(0) & 0.124(0) & 0.319(2) & 0.253(2) & 0.194(1) & 0.534(2) & 0.468(0) & 0.467(0) & 0.379(5) & 0.024(0) & 0.020(0) \\
\rowcolor{gray!20} \cellcolor{white} & None & 0.245(3) & 0.177(4) & 0.123(0) & 0.374(1) & 0.279(1) & 0.193(0) & 0.593(1) & 0.474(0) & 0.469(0) & 0.483(2) & 0.387(3) & 0.022(0) \\
\midrule
\multirow{7}{*}{\rotatebox[origin=c]{90}{\ipm}} & \sys & \textbf{0.392(26)} & \textbf{0.378(34)} & \textbf{0.333(42)} & \textbf{0.469(14)} & \textbf{0.442(19)} & \textbf{0.414(40)} & \textbf{0.689(14)} & \textbf{0.677(36)} & \textbf{0.641(39)} & \textbf{0.545(56)} & \textbf{0.511(57)} & \textbf{0.277(95)} \\
 & \fltrust & \underline{0.317(23)} & \underline{0.293(49)} & 0.199(48) & 0.418(32) & 0.394(26) & \underline{0.345(49)} & 0.645(18) & \underline{0.623(35)} & 0.564(42) & \underline{0.497(5)} & 0.328(134) & 0.220(123) \\
 & \zenops & 0.308(33) & 0.284(37) & \underline{0.225(37)} & \underline{0.429(14)} & \underline{0.394(42)} & 0.325(55) & \underline{0.661(19)} & 0.603(38) & 0.543(40) & 0.470(53) & \underline{0.373(148)} & \underline{0.224(111)} \\
 & GTS & 0.240(52) & 0.242(32) & 0.221(58) & 0.271(41) & 0.312(60) & 0.278(54) & 0.555(50) & 0.524(51) & 0.555(41) & 0.185(102) & 0.148(99) & 0.095(91) \\
 & CS & 0.240(47) & 0.225(46) & 0.208(57) & 0.242(76) & 0.281(60) & 0.254(56) & 0.522(48) & 0.556(40) & \underline{0.573(56)} & 0.092(97) & 0.155(81) & 0.138(81) \\
 & CWTM & 0.228(38) & 0.256(64) & 0.203(32) & 0.215(61) & 0.292(66) & 0.183(94) & 0.547(19) & 0.585(36) & 0.539(16) & 0.125(65) & 0.096(68) & 0.109(59) \\
\rowcolor{gray!20} \cellcolor{white} & None & 0.219(63) & 0.213(54) & 0.194(59) & 0.259(0) & 0.259(0) & 0.259(0) & 0.542(0) & 0.542(0) & 0.542(0) & 0.157(94) & 0.147(97) & 0.102(75) \\
\bottomrule
\end{tabular}%

}

%% file: sections/experiments.tex
\begin{figure}[t]
    \centering
    \includegraphics[width=\linewidth]{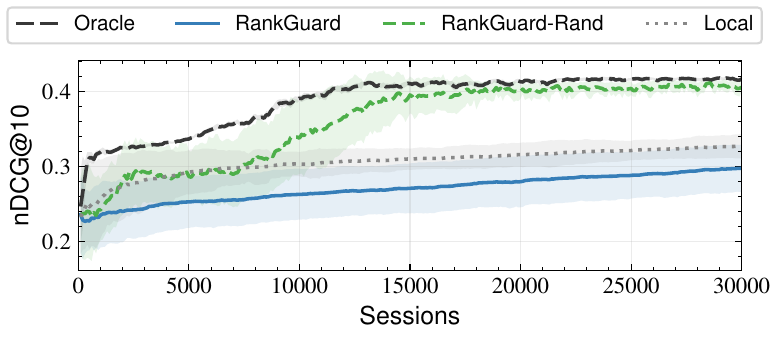}
    \caption{Robustness under the \adapt attack.}
    \label{fig:adaptive}
\end{figure}

\section{Experimental Evaluation}
\label{sec:experiments}

We now present the experimental evaluation of \sys, which answers the following questions:
\begin{enumerate}
    \item What is the robustness of \sys on preserving ranking accuracy compared to defense baselines, for different attacks, click models, and datasets (\Cref{sec:exp_robustness,sec:exp_adapt})?
    \item How do aggregation weights evolve in \sys over time and under different attacks (\Cref{sec:exp_alpha})?
    \item How efficient, in terms of compute time, is \sys compared to baseline defenses (\Cref{sec:exp_efficiency})?
    \item How does the session history size of nodes influence the effectiveness of \sys (\Cref{sec:exp_sessions})?
\end{enumerate}





\subsection{Robustness of \sys}
\label{sec:exp_robustness}

We assess the robustness of \sys from several angles.
We first evaluate \sys against all defense baselines across datasets, click models, and three poisoning attacks (\flip, \lie, \ipm).
Next, we stress-test \sys against \adapt, a worst-case adaptive attack tailored specifically to defeat our defense, and introduce a lightweight modification that neutralizes it.
We then show how the aggregation weight $\alpha$ evolves as honest and malicious models are received. 
We then examine how robustness holds as the share of malicious nodes varies from 0\,\% to 50\,\%. 
Finally, we demonstrate that \sys's robustness generalizes to a neural ranker.

\begin{figure*}[!ht]
    \includegraphics[width=\textwidth]{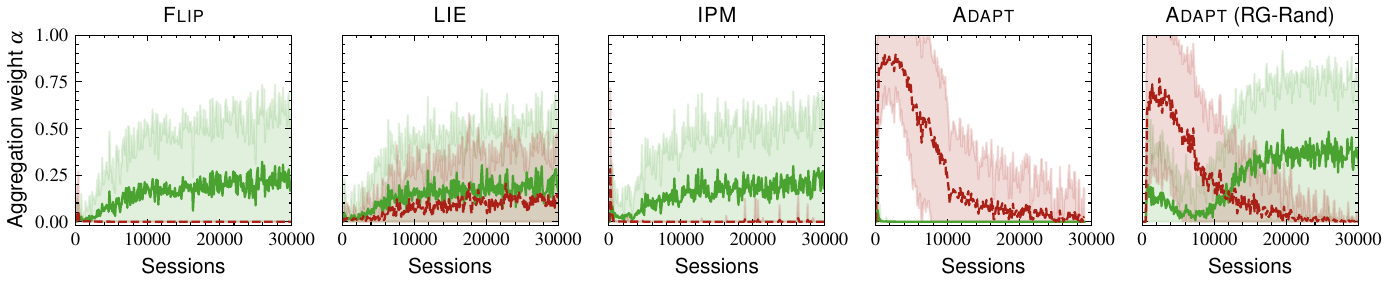}
  \caption{Evolution of $\alpha$ for different attacks from \textcolor{green}{honest} vs. \textcolor{pink}{malicious} users.}
  \label{fig:alpha}
\end{figure*}

\subsubsection{Main benchmark}
\label{sec:exp_main_benchmark}
We run \sys and baselines for \num{30000} sessions, which is sufficient for all systems to converge.
For this experiment, we consider three additional reference baselines:
\begin{enumerate}
    \item \emph{Oracle:} Omniscient nodes always reject models received from malicious nodes and always average models received from honest nodes with weight $\alpha=0.5$.
    \item \emph{Local:} Nodes never exchange models. This is equivalent to training on local click data only, without any collaboration.
    \item \emph{None:} Nodes always average received models (no defense).
\end{enumerate}
\Cref{tab:robustness} reports mean nDCG@10 for all honest nodes after \num{30000} global sessions.\footnote{To reduce evaluation noise, we report nDCG@10 averaged over the last 10 rounds (\num{29000}--\num{30000} sessions) rather than the final round alone.}
To view each baseline's convergence with respect to the number of global sessions, see Appendix~\ref{app:robustness}.
After \num{30000} sessions, each node has processed 300 queries locally.
This can suffice to train an effective ranker with linear \ac{PDGD}~\cite{oosterhuis2018differentiable}, which is why the \emph{Local} baseline remains competitive.
This advantage, however, hinges on the assumption that data remains stationary.
Under concept drift, local interactions no longer suffice, motivating collaborative mechanisms~\cite{li2019cascading}.
As expected from prior work~\cite{baruch2019little,xie2020fall}, statistical defenses (CS, GTS, CWTM) collapse under \lie and \ipm.
\sys, \fltrust, and \zenops, which assess models based on performance rather than statistics, are less affected by these attacks.
Across attacks, datasets, and click models, \textbf{\sys ranks first in the majority of settings} compared to all baselines, sometimes even above \emph{Oracle}.
Note that \emph{Oracle}, while knowing whether a model is received from a malicious node, does not guarantee optimal performance.
Where \sys does not come out first, the gap to the best-performing baseline is usually small.\mv{Again, be specific and concrete - give numbers and specific settings!}
Curiously, one setting stands out as an exception: \emph{informational} on \emph{Istella}, across all attacks. \mg{why}
In summary, our main benchmark results show that \sys consistently delivers robust ranking accuracy across all evaluated settings.

\subsubsection{Adaptive attack.}
\label{sec:exp_adapt}
We designed \adapt as a worst-case attack tailored to defeat \sys (see \Cref{sec:attacks}).
This attack assumes that the attacker knows the victim's session history.
\Cref{fig:adaptive} shows ranking accuracy across sessions for \sys, the \emph{Local} baseline and a modified version of \sys.
Our evaluation confirms that \adapt is effective: the \emph{Local} baseline outperforms \sys, showing that the attacker successfully degrades users' rankers.
To thwart even this powerful attack, we propose a simple modification to \sys in scenarios where the attacker may gain knowledge of users' session history $Q$.
By evaluating each received model on a (freshly sampled) random subset $\tilde{Q}\subset Q$ instead of all of $Q$, \sys becomes non-deterministic: the attacker can no longer compute $\alpha$ exactly, and thus cannot craft an update guaranteed to be accepted.
We evaluate this method across subset sizes and find that \sys performs best when using only 20\,\% of the session history (shown as \sys-Rand. in \Cref{fig:adaptive}).
At \num{17500} sessions, \sys-Rand. exceeds nDCG@10 of $0.4$ and nearly matches \emph{Oracle}, compared to $0.28$ for \sys without random subsampling.
We conclude that this variant of \sys neutralizes the attack once users have accumulated a sufficiently large $Q$.
In this setting, this happens between \num{8000}--\num{15000} sessions (\ie, local history lengths of 80--150).
The evolution of $\alpha$ over this window offers an interesting insight into what is happening.

\subsubsection{Evolution of $\alpha$ in \sys}
\label{sec:exp_alpha}
\Cref{fig:alpha} shows the evolution of aggregation weight $\alpha\in[0,1]$ across sessions for honest and malicious models, across all four attacks.
As the experiment progresses, local session histories grow and models converge, giving \sys increasing certainty in distinguishing honest from malicious updates.
Under \flip and \ipm, \sys achieves near-perfect separation, with malicious models almost always rejected outright ($\alpha\approx 0$).
In contrast, under \lie it is more difficult to distinguish honest from malicious models under the \lie attack.
Nevertheless, \sys consistently aggregates honest models with higher weights and shows robustness in most scenarios (see \Cref{tab:robustness}).

In \Cref{sec:exp_adapt}, we established that a non-deterministic variant of \sys is needed to neutralize the \adapt attack.
In \Cref{fig:alpha}, we compare \adapt on the original \sys with \adapt on \sys-Rand.
Both show the attacker dominating early, when short session histories make \sys's verdict unreliable; the poisoned model's $\alpha$ peaks before declining as histories grow. 
In the former, however, honest models are rejected throughout, so the node never benefits from collaboration even as the attack weakens. 
The latter breaks this: honest $\alpha$ rises steadily and overtakes the poisoned $\alpha$ after around \num{12000} sessions, restoring useful aggregation.

\begin{figure*}[ht]
    \includegraphics[width=\textwidth]{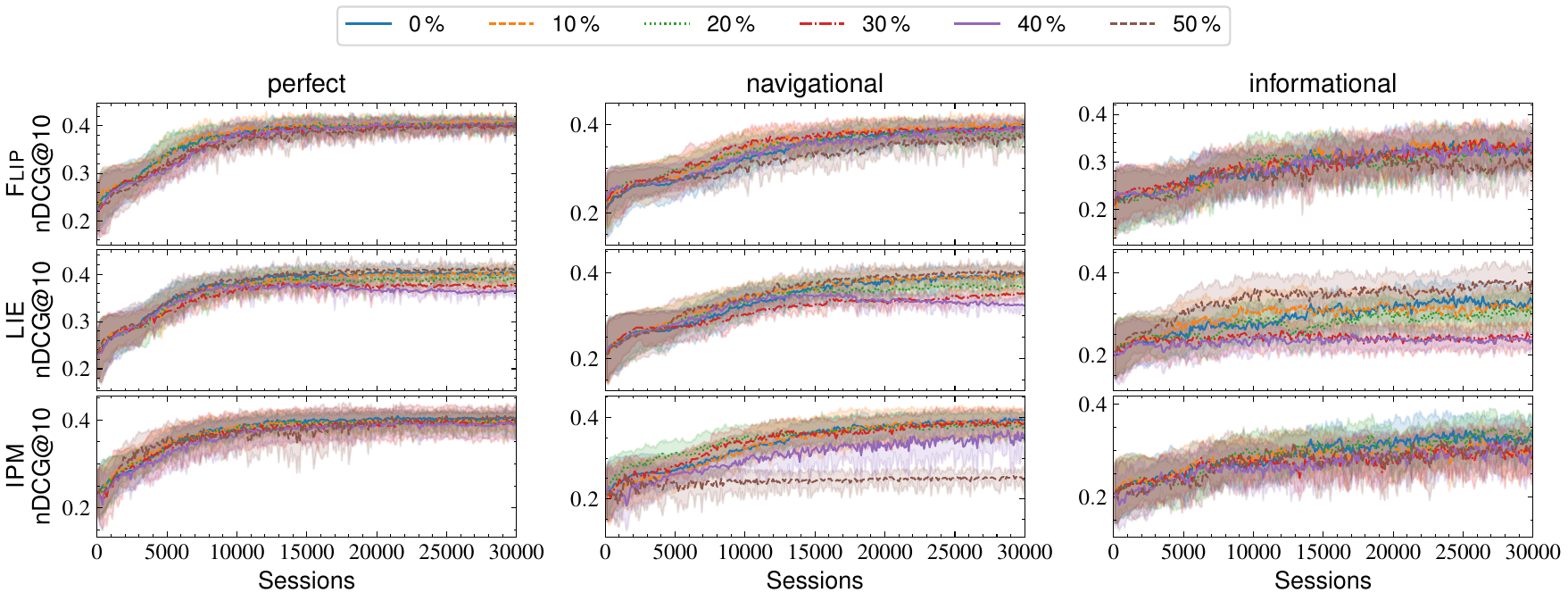}
  \caption{Robustness of \sys under varying shares of malicious nodes in the network.}
  \label{fig:attack-shares}
\end{figure*}

\subsubsection{Varying the attack fraction}
We now study the robustness of \sys varying the attacker fraction $\beta$.
\Cref{fig:attack-shares} show how ranking accuracy evolves for the \flip, \lie and \ipm attack, for three click models and while varying $\beta$ between 0\,\% and 50\,\%.
We do not evaluate $\beta>0.5$, as an adversarial majority breaks the assumptions of the peer sampler and lies beyond the reach of any aggregation defense. \mg{is this argument convincing?}
Note that lower attack shares not only indicate fewer malicious nodes but also more honest nodes, and thus more useful updates to learn from.
These effects cannot be clearly isolated.
Under the \emph{perfect} click model, the curves for different attack shares are nearly indistinguishable, all converging to nDCG@10 $\approx 0.4$.
The \emph{navigational} and \emph{informational} click models introduce more variance and slower convergence, as expected from their noisier feedback.
Particularly difficult to interpret are the results of \lie with the \emph{informational} click model.
Surprisingly, $\beta=0.5$ outperforms $\beta=0$.
We attribute this to \lie's degeneration under round-robin (see \Cref{sec:attacks}): with honest models tightly clustered, the attack reduces to small noise around the honest mean, which under the already-noisy informational signal acts as mild regularization rather than a genuine attack.
Under \ipm with the \emph{navigational} click model, $\beta=0.5$ stalls convergence entirely, plateauing at nDCG@10 $\approx 0.25$ while all lower attack shares reach $\approx 0.4$, indicating the need for an honest majority. The same cannot be observed for \emph{informational}.
We suspect this stems from how \ipm is implemented:
the attacker sends $-\epsilon g$, where $g$ is the honest gradient after completing one search session.
The \emph{navigational} click model produces fewer but more relevance-concentrated clicks, yielding a cleaner, more directed gradient; reversing it thus yields a more directed harmful update.
\mg{conclusion is needed}

\subsubsection{Neural rankers.} 
\begin{figure}[ht]
    \centering
    \includegraphics[width=\linewidth]{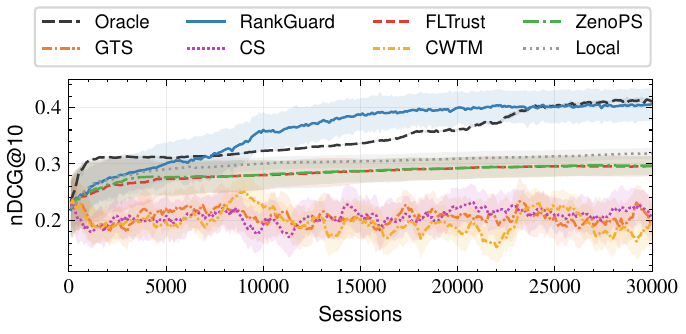}
    \caption{Robustness under \ipm attack using a \emph{neural} ranker.}
    \label{fig:neural}
\end{figure}

To assess whether \sys's robustness holds in the non-convex regime, we repeat our evaluation using a neural ranker. \mg{web30k perfect}
\Cref{fig:neural} shows that \sys continues to outperform the baselines across attacks, indicating that its defense does not depend on convexity and generalizes to non-linear rankers.
It requires roughly 1.5--3$\times$ more sessions until convergence, depending on the attack.
At the point of convergence, it reaches slightly higher accuracy than the linear ranker.
This mirrors centralized experiments of \ac{PDGD}~\cite{oosterhuis2018differentiable}.
In summary, \sys remains effective with neural rankers; the slower convergence and small accuracy gain are properties of the ranker, not of the defense.

\subsection{Efficiency of \sys}
\label{sec:exp_efficiency}

We now quantify the computational efficiency of \sys and baselines.
\Cref{tab:runtime} reports the runtime of \sys and its two most competitive baselines, \fltrust and \zenops, across two ranking models and varying session history sizes, averaged over 100 trials.
Specifically, this corresponds to Step~{7} in \Cref{fig:workflow}.
Note that while population-based filters (CS, GTS, and CWTM) are computationally very cheap, they do not fit the pairwise aggregation setting motivated earlier and, as our results show, are not competitive in robustness. We therefore exclude them from this comparison.
The computational overhead of \sys scales with the number of sessions (or precisely, the number of preference pairs in $\mathcal{P}^{(q)}$) used to evaluate incoming models.
The same is the case for \fltrust and \zenops.
However, \sys is cheaper per session: it only requires forward passes through the model, whereas \fltrust and \zenops replay sessions, involving \ac{PDGD} gradient estimation and update steps.
For example, at 100 sessions \sys requires roughly 130\,ms, versus the 670\,ms needed by \fltrust and \zenops.
This makes \textbf{\sys with a linear ranking model up to 62$\times$ faster than its competitors.}
We additionally evaluate the computational efficiency of the defenses when PDGD uses a neural ranker.
Even here, \sys remains the fastest defense, retaining an $11\times$ speedup over \fltrust and \zenops.
The reduced speedup with a neural ranker indicates that \fltrust and \zenops are dominated by model-agnostic replay overhead, particularly computing \ac{PDGD} bias weights during session replay (\Cref{eq:rho}).
Hence, replacing the linear scorer with a neural scorer adds only a small relative cost to their already expensive pipeline. 
In contrast, \sys reuses cached bias weights and primarily performs model scoring for the defense; neural scoring therefore accounts for a larger share of its runtime, reducing the relative speedup while still leaving \sys substantially faster.
This efficiency is critical, because it lets \sys draw on a larger $Q$ for more reliable judgments, as well as process a higher volume of models in the same time budget.

\begin{table}[t]
\centering
\caption{Runtime (seconds) for the evaluation of one received model in relation to the local session history length.}
\small%
\input{tabs/speed}
\label{tab:runtime}
\end{table}

\subsection{How Many Sessions Are Needed?}
\label{sec:exp_sessions}

\sys's defense ability hinges on the size of the user's session history $Q$.
Larger $|Q|$ lets \sys estimate $X_q$ more reliably, tightening the $t$-statistic and producing more stable $\alpha$ values.
Conversely, smaller $|Q|$ lets sampling noise dominate \sys's verdict on received models, and even honest peers may be rejected while malicious ones slip through.
We now investigate how much history is needed before \sys's decisions stabilize.

\begin{figure}[t]
  \centering
  \includegraphics[width=\linewidth]{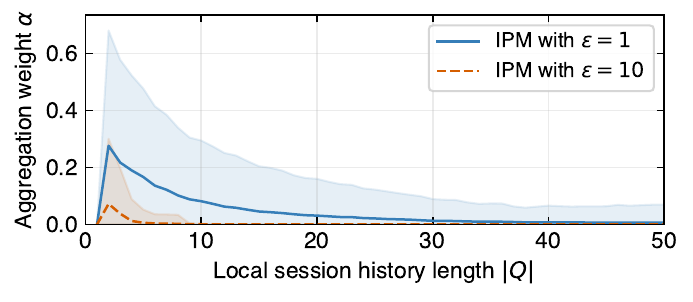}
  \caption{Attack detectability in relation to local session history length measures as the aggregation weight \sys assigns to a malicious model under \ipm.}
  \label{fig:queries}
\end{figure}

To this end, we initialize a model $\theta_u$ randomly and train it on clicks generated from 1000 randomly sampled queries using the \emph{perfect} click model.
This gives a fairly converged ranker, as well as a local history with $|Q_u|=1000$.
We then construct a poisoned model $\theta_v$ by computing the gradient on a copy of $\theta_u$ for clicks generated from a freshly sampled query, and updating the copy in the opposite (\ie, harmful) direction.
Note that this is equivalent to the \ipm attack with $\epsilon=1$ (our previous experiments used $\epsilon=10$), and represents a minimal and harder to detect attack.
By construction, $\theta_v$ is strictly worse than $\theta_u$ at explaining the user's clicks: it was derived from $\theta_u$ by stepping away from the honest gradient. The correct verdict from \sys is therefore full rejection, $\alpha=0$. Any $\alpha>0$ means the poisoned model has been (partially) incorporated; the gap from zero quantifies the defense's failure on this minimal attack.
We evaluate the aggregation weight $\alpha$ that \sys assigns to the poisoned model as a function of the suffix length of $Q_u$ used to compute it.
We repeat this experiment over 1000 seeds and report the mean and standard deviation in \Cref{fig:queries}.
Even in the scenario where the attacker deviates from the victim model minimally ($\epsilon=1$), \sys rejects the poisoned model ($\alpha\approx 0$) after only around 40 sessions; for the stronger $\epsilon=10$ from our main experiment, the poisoned model is \emph{always} rejected after only 10 local sessions.
In short, \sys needs remarkably little history to defend effectively, allowing new participants to benefit from collaboration early.
This responsiveness also matters under concept drift: as a user's interests shift, \sys can assess incoming models against only the user's most recent sessions.

%% file: tabs/speed.tex
\begin{tabular}{llrrr}
\toprule
Defense & Model & $|Q|=10^2$ & $|Q|=10^3$ & $|Q|=10^4$ \\
\midrule
\sys & \emph{Linear} & 0.012 (0.001) & 0.129 (0.003) & 1.448 (0.032) \\
 & \emph{Neural} & 0.073 (0.013) & 0.781 (0.145) & 8.144 (1.151) \\
\fltrust & \emph{Linear} & 0.669 (0.127) & 8.057 (0.487) & 92.102 (1.286) \\
 & \emph{Neural} & 0.721 (0.138) & 8.290 (0.821) & 94.963 (1.708) \\
\zenops & \emph{Linear} & 0.668 (0.127) & 8.058 (0.487) & 92.144 (1.277) \\
 & \emph{Neural} & 0.721 (0.139) & 8.164 (0.537) & 94.688 (1.393) \\
\bottomrule
\end{tabular}

%% file: sections/related_work.tex
\section{Related Work}

In \ac{FL}, robust aggregation techniques such as \textsc{Krum}~\cite{blanchard2017machine}, \textsc{CWTM}~\cite{yin2018byzantine}, and \textsc{Geometric Median}~\cite{pillutla2022robust} assume an honest majority and treat malicious updates as statistical outliers.
While effective against naive attacks, they break under carefully crafted malicious updates that remain within the honest distribution~\cite{baruch2019little,xie2020fall,fang2020local}.
Many works~\cite{yang2019byrdie,fang2022bridge,he2022byzantine} adapt robust aggregation rules such as \textsc{Krum} and coordinate-wise trimming to \ac{DL} but suffer from the same weaknesses.
This motivated \emph{trusted-reference} defenses that score updates by their effect on a benign validation dataset.
In \ac{FL}, \fltrust~\cite{fltrust} compares each client update to a server-side reference update by cosine similarity, while \textsc{Zeno} and its variants~\cite{xie2020zeno++,xie2022zenops} rank updates by their estimated loss reduction on a small clean dataset held by the server. 
In \ac{DL}, \textsc{UBAR}~\cite{guo2021byzantine} removes the server dependence by having nodes screen model updates with their local data, and combines a distance-based pre-selection with a loss-based acceptance test that keeps only updates improving the node's local objective.
\textsc{Sentinel}~\cite{feng2024sentinel} extends this idea with a hybrid of similarity filtering, local-loss validation, and norm normalization.
\textsc{Basil}~\cite{elkordy2022basil} also applies the local-validation principle but processes models over a logical ring with a memory of past updates.
All these methods assume synchronous, fixed-topology communication and a pool of simultaneously available neighbor models.
Yet, many of the assumptions break under the asynchronous, push-based, churn-prone dynamics of gossip learning we consider in this work.
The principle of loss-based evaluation is closest to our approach in spirit, but they do not directly apply to \ac{OLTR}: \ac{PDGD}~\cite{oosterhuis2018differentiable} ascends an unbiased gradient estimate rather than minimizing an explicit loss.
Furthermore, nodes only have click data available, which is noisy and biased.

The \ac{OLTR} setting has remained understudied by the Byzantine-robust learning community.
The only systematic study, by \citet{wang2023analysis}, evaluates untargeted poisoning attacks with robust aggregators for federated \ac{OLTR}, finding that these substantially harm ranking quality.
We close this gap with \sys, a gossip-native defense for decentralized \ac{OLTR}.
A parallel line of work~\cite{gregoriadis2025decentralized,gregoriadis2025swarmsearch} approaches decentralized \ac{LTR} by gossiping raw click data rather than model weights.
\textsc{SwarmSearch}~\cite{gregoriadis2025swarmsearch} employs Data Shapley~\cite{ghorbani2019data} to value peers' data and discard poisoned contributions. 
This paradigm is ill-suited to our setting: Data Shapley requires expensive Monte Carlo retraining; sharing raw interaction data sacrifices privacy, and data gossip is communication-inefficient compared to model gossip.
Moreover, being batch-based rather than \emph{online} \ac{LTR}, it does not learn from live, position-biased clicks.

%% file: sections/conclusion.tex
\section{Conclusion}

We presented \sys, a decentralized framework for \ac{OLTR} in which users collaboratively train ranking models through peer-to-peer gossip, without any central authority.
Rather than filtering out statistical outliers among peer updates, each node judges incoming models by how well they explain its own position-bias-corrected click history, a design we support with the first formal convergence analysis of a decentralized OLTR algorithm. 
Our evaluation shows this approach is both effective and practical: 
\sys delivers the strongest ranking accuracy across most attacks, datasets, and click models, while running up to 62$\times$ faster than comparable defenses, making it efficient enough for Internet-scale deployment. 
It also defends reliably from very short click histories, so new participants benefit from collaboration almost immediately, with the defense adapting gracefully as user interests drift over time.

%% file: appendix/notations.tex
\section{List of Notation}\label{app:notation}

\begin{tabularx}{\columnwidth}{@{}lX@{}}
\toprule
\textbf{Symbol} & \textbf{Description} \\
\midrule
$q$ & Session identifier \\
$D^{(q)}$ & Candidate document set for query in session $q$ \\
$\bm{d}$ & Feature vector of a query-document pair \\
$d$ & Dimensionality of $\bm{d}\in\mathbb{R}^d$ \\
$f_\theta(\cdot)$ & Scoring function with parameters $\theta$ \\
$\theta$ & Model weights \\
$R$ & Sampled ranking of length $k$ \\
$R^*$ & Ranking $R$ with one click pair $\bm{d_i} >_c \bm{d_j}$ swapped \\
$\bm{d_i} \succ \bm{d_j}$ & pairwise preference indicating that document $\bm{d_i}$ should be ranked above $\bm{d_j}$ \\
$\bm{d_i} >_c \bm{d_j}$ & Click-inferred preference: $\bm{d_i}$ was clicked and $\bm{d_j}$ was examined but not clicked \\
$\mathcal{P}^{(q)}$ & Set of pairwise click preferences inferred for session $q$ \\
$\rho^{(q)}_{ij}$ & Position-bias weight of pref. pair $\bm{d_i} >_c \bm{d_j}$ in session $q$ \\
$\eta$ & Learning rate \\
$n$ & Number of nodes in the network \\
$\beta$ & Fraction of malicious nodes in the network \\
$u, v$ & Node identifiers (in pairwise settings we denote $u$ as the local node and $v$ as a peer) \\
$\theta_u$ & Model weights of node $u$ \\
$Q_u$ & Session history of node $u$ \\
$S(q,\theta)$ & Bias-corrected session performance score \\
$\Delta S$ & Mean score improvement of received model against local model \\
$X_q$ & Per-query score difference $S(q,\theta_v)-S(q,\theta_u)$ \\
$\alpha$ & Aggregation (interpolation) weight, $\sigma(\kappa t)$ \\
$\tilde{Q}\subset Q$ & Random subset of $Q$ (anti-adaptive variant) \\
\bottomrule
\end{tabularx}

%% file: appendix/convergence_proof.tex
\section{Proof Sketch of Theorem~\ref{th:convergence}}\label{app:convergence_proof}

Under the assumptions and formalism of \sys's workflow presented in \Cref{sec:theory}, we reduce our system down to a weighted aggregation variant of \elloc~\cite{de2023epidemic} with an in-degree of $1$ for each local round of every node in the network. Recall that $(i)$ $\hat{\alpha}$ and $\sigma_{\hat{\alpha}}^2$, respectively, are the mean and variance of the weight any node assigns, in any of its local rounds, to the received model, and $(ii)$ node $u$, in any local round $t_u$, assigns a weight of $1 - \alpha_{uv}^{(t_u)}$ to its own model and $\alpha_{uv}^{(t_u)}$ to the received model. 

Incorporating this weighted aggregation scheme, we determine the expected variance, $V$, of the mixing matrix by 
$$V = \mathbb{E}\left[\left(1 - \alpha_{uv}^{(t_u)}\right)^2 + \left(\alpha_{uv}^{(t_u)}\right)^2\right] = \left(1 - \hat{\alpha}\right)^2 + \hat{\alpha}^2 + 2\sigma_{\hat{\alpha}}^2.$$
On the other hand, from the convergence analysis of \elloc, we have $V = \frac{1}{s+1}$ for any generic in-degree $s$. Combining these, we get the expected and effective in-degree $s^*$ for each node in each of its local rounds as 
$s^* = \frac{1}{V} - 1 = \frac{2\hat{\alpha}(1 - \hat{\alpha}) - 2\sigma_\alpha^2}{(1 - \hat{\alpha})^2 + \hat{\alpha}^2 + 2\sigma_{\hat{\alpha}}^2}.$
Plugging this expected and effective in-degree into the proof of Theorem 1.b) of \elloc derives the convergence bound for \sys.


%% file: appendix/robustness_plots.tex
\onecolumn
\section{Robustness Plots}\label{app:robustness}

\begin{figure*}[ht]
  \centering
  \includegraphics[width=\textwidth]{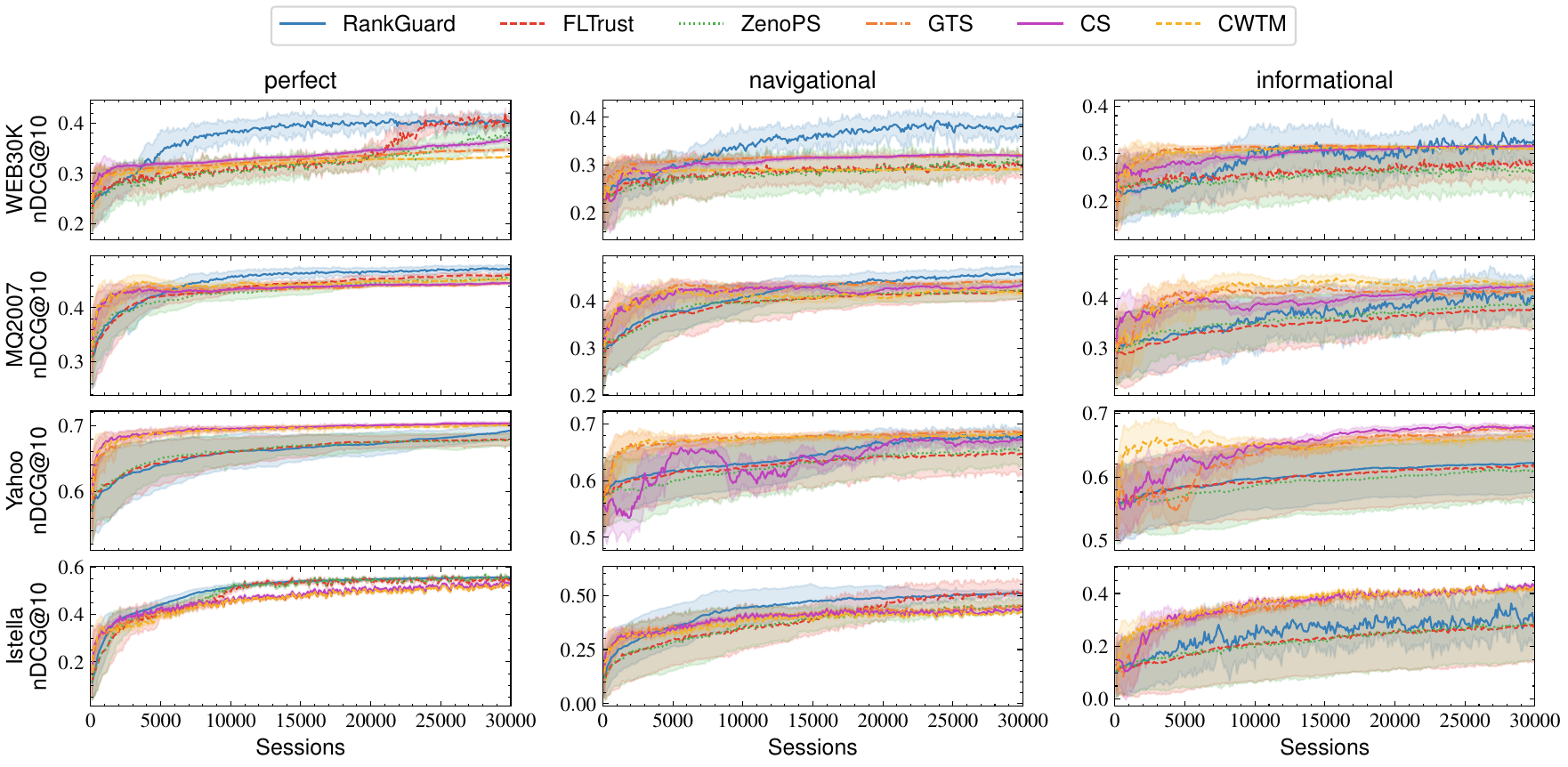}
  \caption{Robustness under the \flip attack with attack share $\beta=0.2$.}
  \label{fig:flip}
\end{figure*}
\begin{figure*}[ht]
  \centering
  \includegraphics[width=\textwidth]{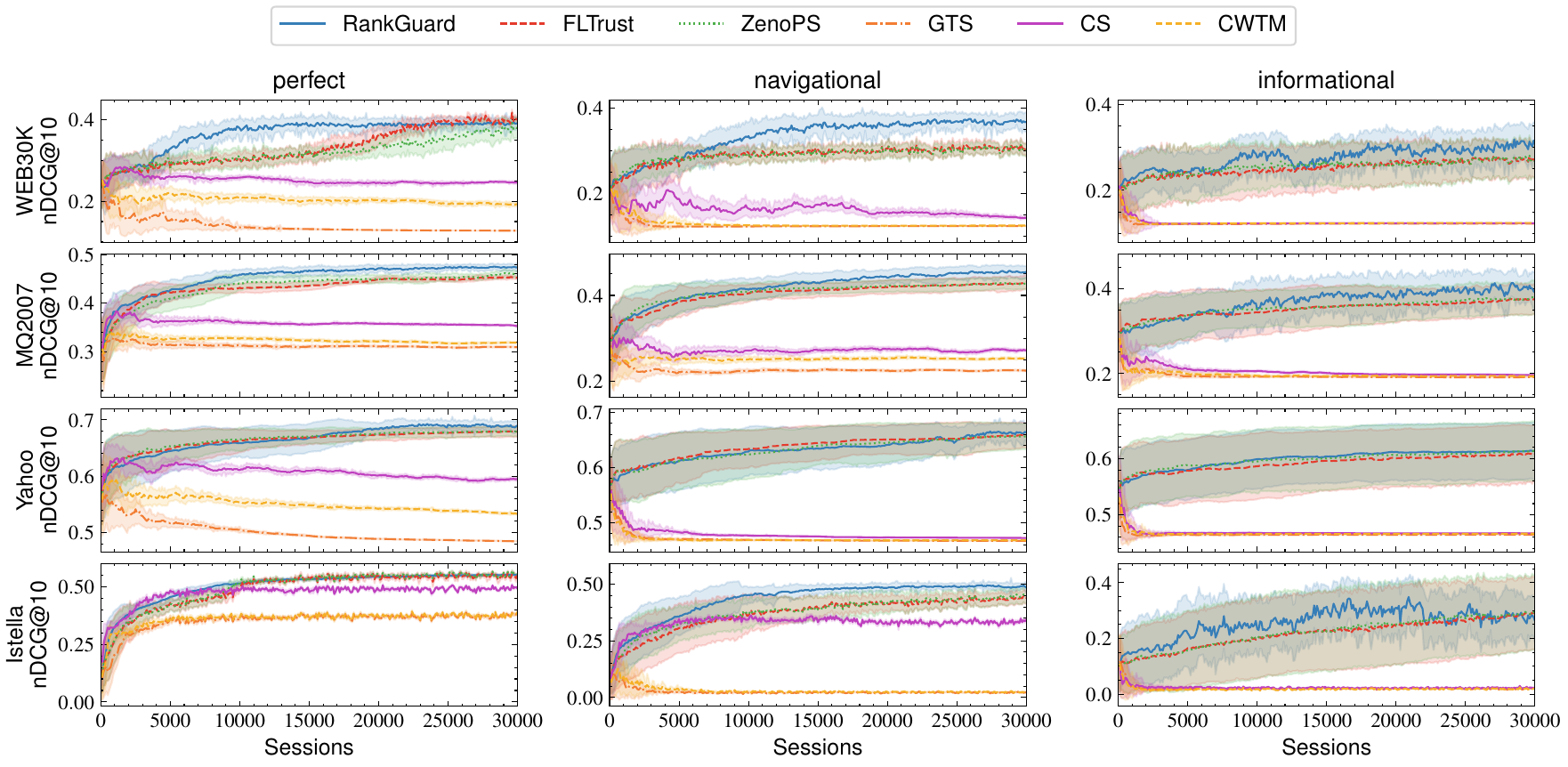}
  \caption{Robustness under the \lie attack with attack share $\beta=0.2$.}
  \label{fig:lie}
\end{figure*}
\begin{figure*}[ht]
  \centering
  \includegraphics[width=\textwidth]{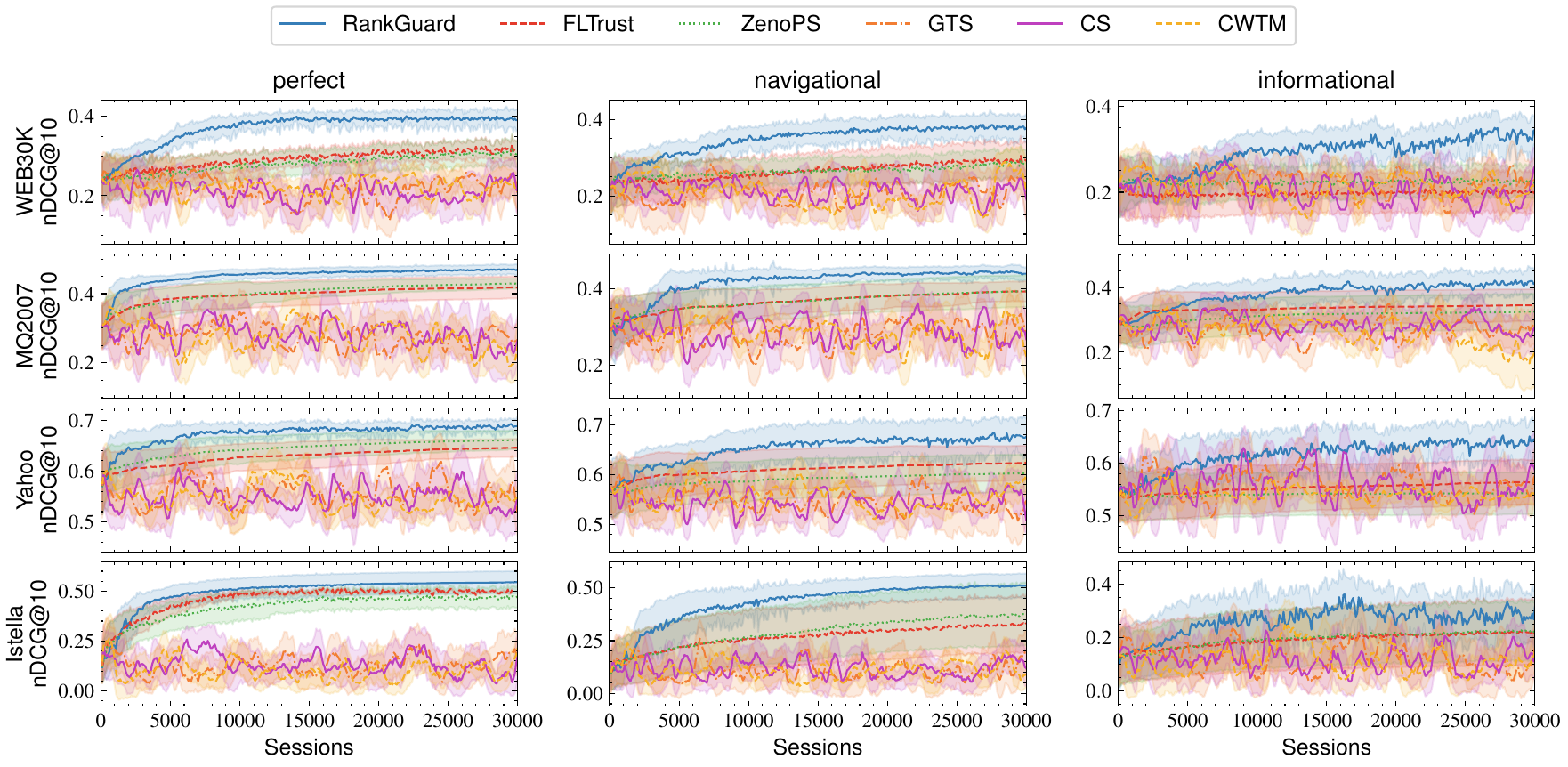}
  \caption{Robustness under the \ipm attack with attack share $\beta=0.2$.}
  \label{fig:ipm}
\end{figure*}
\twocolumn